\documentclass[lettersize,journal]{IEEEtran}
\usepackage{amsmath,amsfonts}
\usepackage{algorithm}
\usepackage{algpseudocode}  
\usepackage{array}
\usepackage[caption=false,font=normalsize,labelfont=sf,textfont=sf]{subfig}
\usepackage{textcomp}
\usepackage{stfloats}
\usepackage{url}
\usepackage{xurl}
\usepackage{verbatim}
\usepackage{graphicx}
\usepackage{cite}
\usepackage{balance}
\usepackage{amssymb}

\begin{document}

\title{Bit-Level Triangular Content-Aware Permutation for Fragile Image Watermarking: Zero False Positive Rate, Single-Bit Sensitivity, and Arbitrary Dimension Support}

\author{Zahra Ghoraeian,
        Mohammad-Reza Sadeghi,
        and Samaneh Mashhadi,
\thanks{Z. Ghoraeian is with the Department of Mathematics and Computer Science, Amirkabir University of Technology (Tehran Polytechnic), Tehran, Iran (e-mail: zahra.ghoraeian@aut.ac.ir). ORCID: 0009-0009-3547-4437.}
\thanks{M.-R. Sadeghi is with the Department of Mathematics and Computer Science, Amirkabir University of Technology (Tehran Polytechnic), Tehran, Iran (e-mail: msadeghi@aut.ac.ir). ORCID: 0000-0002-7676-4168.}
\thanks{S. Mashhadi is with the School of Mathematics and Computer Science, Iran University of Science and Technology, Narmak, Tehran, 1684613114, Iran (e-mail: smashhadi@iust.ac.ir). ORCID: 0000-0001-9191-1376.}
\thanks{Corresponding author: Mohammad-Reza Sadeghi (e-mail: msadeghi@aut.ac.ir).}
}
\markboth{Arxiv}%
{Ghoraeian \MakeLowercase{\textit{et al.}}: Bit-Level Triangular Content-Aware Permutation for Fragile Image Watermarking}

\maketitle

\begin{abstract}
With the growth of digital document exchange, protecting image integrity against attacks such as Vector Quantization (VQ) and collage has become critical. Existing methods are not only vulnerable to these attacks but also limited to fixed image dimensions. This paper presents a novel, dimension-agnostic, single-phase fragile watermarking algorithm that enhances both security and tamper localization accuracy by replacing conventional hash functions with Triangular Content-Aware Permutation (TCA).

In the proposed scheme, the image is first combined with a key-based global noise and then divided into $6\times 6$ blocks. The core innovation lies in applying a content-dependent permutation with intrinsic avalanche effect (TCA) at the bit-plane level, which, together with a block-location-dependent coefficient, generates a unique content-dependent watermark. For color images, instead of independent channel processing, a vertical sandwich transformation is proposed that merges the three channels at the block level, preserves inter-channel dependency, and increases execution time by approximately 1.7 times. The "remainder merging" strategy eliminates dimension constraints, allowing the algorithm to run without padding. The proposed method works on both color and grayscale images with different bit-depths and square/rectangular blocks, enabling users to trade off speed versus localization granularity.

Experimental evaluation on 50 grayscale and 10 color images under 18 different attacks shows that the method achieves FPR = 0\% and FNR = 0\% for 17 attacks. Only for salt-and-pepper noise does the method deviate slightly from zero while remaining negligible, with an average FNR of 0.27\% for grayscale and 0.14\% for color images. Visual transparency is confirmed by average PSNR values of 51.14 dB (8-bit), 75.25 dB (12-bit), and 99.33 dB (16-bit). The average embedding and extraction times for grayscale images are 1.61 s and 1.63 s, respectively.

The algorithm achieves 100\% accuracy against collage, VQ, copy-move, JPEG compression (quality 5–95), and geometric attacks. By replacing static mathematical formulas with dynamic content-dependent geometry, the proposed scheme provides a secure and flexible solution for sensitive applications such as digital forensics, medical image archiving, and legal document authentication.
\end{abstract}

\begin{IEEEkeywords}
Fragile watermarking, Content-Aware Permutation (TCA), tamper localization, image authentication, bit-level security, digital forensics.
\end{IEEEkeywords}

\section{Introduction}

\IEEEPARstart{W}{ith} the digitalization of assets, concepts such as digital currency, telemedicine, and the Internet of Things have emerged. Digital data is transmitted through insecure public channels, jeopardizing security. Images in sensitive domains such as forensics, medical imaging, and military systems are considered legal and strategic documents; therefore, any unauthorized modification has irreparable consequences. In this context, fragile watermarking offers an effective solution for ensuring integrity and precisely localizing tampering.

\subsection{Background}
With the digitalization of assets, concepts such as digital currency, telemedicine, and the Internet of Things have emerged. Digital data is transmitted through insecure public channels, jeopardizing security. Images in sensitive domains such as forensics, medical imaging, and military systems are considered legal and strategic documents; therefore, any unauthorized modification has irreparable consequences. In this context, fragile watermarking offers an effective solution for ensuring integrity and precisely localizing tampering~\cite{Sreenivas2017,Amrullah2025,Sharma2024}.

\subsection{Main Challenges of Fragile Watermarking}
Fully fragile watermarking is sensitive to any content modification~\cite{Sreenivas2017}. The first challenge is its block-wise nature~\cite{Amrullah2025}. The block size must strike a balance between localization accuracy and embedding capacity; larger blocks reduce accuracy~\cite{Sreenivas2017,Sharma2024}. The second challenge is resistance to advanced attacks such as Vector Quantization (VQ)~\cite{Sharma2024,Sreenivas2017}, copy-move~\cite{Sreenivas2017,Liu2025}, and collage~\cite{Sreenivas2017,Dhole2015}. To counter these, the watermark of each block must depend on both the block content and its position~\cite{Sharma2024}. The third challenge is the ability to generalize to other color spaces and different bit-depths; many algorithms are designed solely for 8-bit grayscale images~\cite{Sreenivas2017} and cannot be extended to color images or high bit-depths (12/16 bits used in medical imaging and remote sensing)~\cite{Priyadarshini2026,Bouarroudj2025}. The fourth challenge concerns JPEG compression behavior and bit-level sensitivity. A fragile watermark must be sensitive to any change, including compression, since repeated JPEG compression can indicate tampering~\cite{Liu2025}. Furthermore, bit-level sensitivity analysis has been neglected in most works.

\subsection{Research Gaps}
Despite extensive efforts, the following research gaps remain:
\begin{enumerate}
\item \textbf{Non-zero false positive rate:} many methods cannot guarantee zero FPR, and even advanced methods achieve zero FPR only ``in most cases''~\cite{Sharma2024}.
\item \textbf{Padding issue and fixed dimension dependency:} many methods assume that the image dimensions are divisible by the chosen block size~\cite{Cedillo2026,Adi2025}.
\item \textbf{Inability for users to freely choose the block size:} in existing methods, the block size is fixed and pre-determined (e.g.,~\cite{Adi2025}), and changing it disrupts the underlying mathematical framework.
\item \textbf{Lack of bit-level sensitivity analysis:} it remains unclear whether a single-bit alteration is detectable.
\end{enumerate}

\subsection{Innovation, Objectives, and Design Principles of the Proposed Method}
To address these gaps, the proposed method is designed with the following key objectives: zero FPR, user-selectable block size, solving the padding problem without adding extra information, generalizability to color spaces and different bit-depths, and full fragility against JPEG compression. Although these objectives may seem ambitious, they all naturally arise from a fundamental paradigm shift from analytic/statistical watermark generation to a content-dependent pure permutation at the bit-plane level.

The core innovation is the use of a pure pixel permutation called ``Triangular Content-Aware Permutation (TCA),'' which has three key properties:
\begin{enumerate}
\item complete content dependency,
\item strong avalanche effect, and
\item independence from input dimensions~\cite{Ghoraeian2026}.
\end{enumerate}

Our innovation is the transfer of TCA from the pixel space to the bit-plane level. This paradigm offers five advantages:

\begin{enumerate}
\item \textbf{Bit-level security and avalanche sensitivity:} any single-bit change transforms the watermark. The method is fully fragile against JPEG.
\item \textbf{Simultaneous content and location dependency:} block content is transformed into a watermark via TCA, and seed-based noise provides location dependency.
\item \textbf{Freedom in block size selection and solving the padding problem:} TCA is inherently independent of dimensions. The padding problem is solved through the ``remainder merging'' strategy.
\item \textbf{Inherent generalizability:} the process is defined in the spatial domain and can be easily extended to color images and different bit-depths (8, 12, 16 bits).
\item \textbf{Guaranteed zero FPR and near-zero FNR:} detection is defined as a deterministic function. Pure permutations like TCA are inherently collision-free (the only potential source of collisions is output truncation). Consequently, FPR is always zero. FNR is also zero under normal conditions (only for salt-and-pepper noise have negligible values below 1\% been observed).
\end{enumerate}

The proposed method has been evaluated against 18 attacks, and the results confirm absolute zero FPR and near-zero FNR.

\subsection{Contributions of the Present Research}
\begin{enumerate}
\item Pure permutation with inherent avalanche effect at the bit level
\item Three-layer independent security architecture (seed-based noise, CTR multipliers, TCA at bit-level)
\item Zero FPR and near-zero FNR (FNR = 0 for 17 out of 18 attacks)
\item User-selectable block size
\item Support for arbitrary dimensions without padding (``remainder merging'' strategy)
\item Empirical verification of watermark uniqueness (zero duplicate watermarks across over 72,820 blocks)
\item Full fragility against JPEG
\item Resistance to advanced attacks (VQ, collage, single-bit change detection)
\item Inherent generalizability (independent of color layers and bit-depth)
\item Imperceptibility (average PSNR of 51.14 dB and SSIM of 0.9975)
\end{enumerate}

\subsection{Paper Organization}
Section 2 covers the theoretical background and prerequisites. Section 3 reviews related work. Section 4 describes in detail the proposed bit-level TCA-based embedding and extraction algorithms. Section 5 presents experimental results, security analysis, and comparisons. Section 6 discusses limitations and future work, and Section 7 concludes the paper.

\section{Theoretical Background and Preliminaries}

\subsection{Definition and Characteristics of Fragile Watermarking}
Fragile watermarking is a branch of digital data hiding aimed at ensuring image integrity and authentication~\cite{Sreenivas2017}. Unlike robust watermarking, a fragile watermark must be destroyed by any modification (even a single pixel or bit). An ideal system has four characteristics~\cite{Sharma2024}: imperceptibility, high fragility, tamper localization capability, and security.

\subsection{Block Partitioning Structure and Its Challenges}
Most methods use a block-wise approach~\cite{Sharma2024}. Choosing the block size involves a trade-off between localization accuracy and embedding capacity: small blocks (e.g., $4\times 4$) offer high accuracy but low capacity, while large blocks (e.g., $16\times 16$) offer the opposite. The padding issue, which arises when image dimensions are not divisible by the block size, remains one of the less-discussed challenges.

\subsection{General Steps of a Fragile Watermarking System}
A typical system consists of five steps~\cite{Sreenivas2017,Sharma2024}:
\begin{enumerate}
\item watermark generation from block content (and possibly its position),
\item watermark embedding (usually in the LSB),
\item watermark extraction from the suspected image,
\item watermark regeneration from the received image, and
\item comparison and tamper map generation.
\end{enumerate}

\subsection{Threats and Attacks in Fragile Watermarking}
According to~\cite{Sharma2024} and~\cite{Liu2025}, the most important attacks include: image processing attacks (JPEG, filtering, noise), block relocation attacks (copy-move, collage), and structural attacks (Vector Quantization — VQ). To counter these, the watermark of each block must depend on both its content and its location.

\subsection{Fundamental Concepts at the Bit Level}
The proposed method operates on the bit-planes of the image. In an image with bit-depth $d$, each pixel is represented as $d$ bits. The $k$-th bit-plane ($k=0$ for LSB, $k=d-1$ for MSB) is a matrix of bits. Bit-level sensitivity refers to the ability to detect single-bit alterations, which has been neglected in many previous methods~\cite{Liu2025}.

\subsection{Cryptographic Hash Functions and Their Application in Fragile Watermarking}
Many advanced methods use cryptographic hash functions (e.g., SHA-512), which have three main properties: pre-image resistance, second pre-image resistance, and collision resistance~\cite{Stinson2019}. In an ideal scenario, if the watermark equals the hash of the block, FPR is zero. However, the main problem is output length (512 bits for SHA-512), which requires a minimum block size of $23\times 23$, reducing localization accuracy~\cite{Sharma2024}. Existing methods use shortened hashes, but this increases collision probability and creates vulnerability to codebook attacks~\cite{Stinson2019}.

\subsection{Triangular Content-Aware Scrambling (TCA) and Pure Permutations}
Unlike hash functions that inherently have collision probability, pure permutations are bijective (one-to-one and onto) and create no collisions. The Triangular Content-Aware Scrambling (TCA) algorithm, introduced in our previous work~\cite{Ghoraeian2026}, is precisely a pure permutation based on image content.

The execution mechanism of TCA is as follows: first, key points of the image (such as edge points) are extracted using the Canny algorithm. Then, the image plane is partitioned into a set of non-overlapping triangles using Delaunay triangulation. The order and arrangement of these triangles, which depend entirely on the distribution of key points and thus on the image content, serve as the permutation map for rearranging image pixels~\cite{Ghoraeian2026}.

The key properties of TCA that distinguish it for authentication applications are:
\begin{enumerate}
\item \textbf{Complete content dependency:} any change in image content alters the triangulation structure and completely changes the permutation map.
\item \textbf{Strong avalanche effect:} a small change in the input produces widespread and unpredictable changes in the output.
\item \textbf{Independence from input dimensions:} the algorithm can be applied to images of any dimensions without special preprocessing.
\item \textbf{Non-analytic nature:} its non-analytic, point-distribution-dependent nature makes it resistant to Known-Plaintext Attacks (KPA).
\end{enumerate}

The only potential source of collisions in TCA is optional output truncation; the permutation itself is inherently collision-free. This fundamental property is the starting point of the present research: in this paper, for the first time, we extend the TCA algorithm from the pixel domain to the bit-plane level, preserving its unique security properties (including the avalanche effect) while enabling single-bit change detection, full fragility against JPEG, and user-selectable block size.

\subsection{The Challenge of Balancing Transparency and False Negative Rate}
A fundamental point: we always use a small number of bits to verify the integrity of a large number of bits. Therefore, collisions are inevitable according to the Pigeonhole Principle~\cite{Stinson2019}. The value of a method lies in keeping the practical FNR as close to zero as possible and ensuring that collisions are not exploitable. Shortened hash-based methods typically suffer from high FNR and sometimes use threshold-based comparison to compensate, which takes FPR away from zero — unacceptable for sensitive applications (medical, military). Transform-domain methods may also suffer from non-zero FPR due to rounding after inverse transform.

\subsection{Evaluation Metrics}
For quantitative evaluation, standard metrics are defined based on comparison between the ground truth tamper map and the detected map~\cite{Sharma2024}:
\begin{itemize}
\item True Positive Rate (TPR)
\item False Positive Rate (FPR) — which must be zero for sensitive applications
\item False Negative Rate (FNR)
\item Accuracy (ACC)
\item Precision
\item F1-Score
\end{itemize}

In an ideal system: TPR = 100\%, FPR = 0\%, FNR = 0\%, ACC = 100\%, Precision = 100\%, F1 = 1~\cite{Sharma2024}.

\subsection{Summary of Foundations}
The key challenges addressed by the proposed method are: non-zero FPR in existing methods, high FNR in shortened hash-based methods, fixed dimension dependency and padding issues, inability for users to freely choose block size, and lack of bit-level sensitivity analysis. In the proposed method (Section 4), transferring TCA to the bit-plane level~\cite{Ghoraeian2026}, together with the intrinsic avalanche effect and a three-layer security architecture, enables achieving absolute zero FPR, single-bit change detection, full fragility against JPEG, freedom in block size selection, and solving the padding problem.

\section{Literature Review}

Fragile watermarking is one of the effective methods for ensuring the integrity and authentication of digital images. By embedding information into the image that is sensitive to any modification, this method enables tamper detection. In recent years, various methods have been proposed to improve the accuracy, security, and efficiency of this field. In the following, a review of methods proposed in 19 recent papers (2022–2026) is presented, and then 7 of these are selected as representatives, analyzed in detail, and compared with the proposed scheme.

\subsection{Introduction to Proposed Methods}
The reviewed methods in this field can be categorized from several perspectives: transform type, embedding approach, and recovery capability.

The first group comprises self-embedding methods in the spatial domain. For instance, Singh et al.~\cite{Singh2023b} divide the image into $2\times 2$ blocks, extract 10 recovery bits from 5 MSBs after applying DCT and 2 authentication bits from position and secret keys, and embed recovery information in two partner blocks; detection is performed at three levels. In a similar approach but using modular arithmetic, Kosuru et al.~\cite{Kosuru2023} in the spatial domain use the average of 4 MSBs and combine it with a logistic map (via XOR) to generate 4 recovery bits, embedding them using a combination of LSB substitution and modular functions—offering computational simplicity as a key advantage. Wu et al.~\cite{Wu2026} propose a self-recovery approach by constructing authentication watermarks from a combination of rotation-invariant LBP, DCT coefficients, and GLCM features, with stepwise embedding using OPAP and LSB. However, the use of thresholds in detection eliminates bit-level sensitivity and ignores changes up to 28 units.

On the other hand, transform-domain methods are also prevalent. Azizoglu and Toprak~\cite{Azizoglu2023} present a fragile and reversible approach in the DWT domain with DE optimization, designed specifically to correct rounding errors. They divide the image into $8\times 8$ blocks and propose two strategies, WBE and MD5, for generating authentication bits. Singh et al.~\cite{Singh2023a} similarly use IWT, dividing the image into $2\times 2$ blocks and generating 10 recovery bits from IWT coefficients and 2 authentication bits from the mean, position, and secret keys, offering immunity to rounding errors. In the DWT domain, Li et al.~\cite{Li2024} generate two separate watermarks (authentication and feature), encrypt them using a chaotic map, embed them in the LSBs of wavelet coefficients, and perform detection at three levels with recovery combining both watermarks.

Some methods specifically focus on high capacity or complete reversibility. Akhtarkavan et al.~\cite{Akhtarkavan2023} in the IIDWT domain with A5 lattice vector quantization provide high capacity for embedding patient metadata, but the use of a fixed watermark introduces security vulnerabilities. In contrast, Bouarroudj et al.~\cite{Bouarroudj2024a} in the DFT domain generate a watermark based on DCT and encrypt it using a Fibonacci Q-matrix, achieving PSNR as high as 117 dB and capacity of 0.25 BPP; however, their claimed reversibility faces rounding error challenges. The same group in another study~\cite{Bouarroudj2024b} increased capacity to 3 BPP and added medical data hiding capability by combining patient data with the Chen chaotic system. Additionally, Bouarroudj et al.~\cite{Bouarroudj2025} in a hybrid DWT-DFT approach proposed a fully blind and fragile method with precise localization (up to 0.005\%) but still suffers from rounding errors. In the domain of high-resolution color images, Cedillo-Hernandez et al.~\cite{Cedillo2026} employ a semi-fragile approach in DCT with QIM-DM and leverage DnCNN and VDSR networks to improve recovery quality, but sacrifice full pixel-level fragility.

In the domain of self-recovery methods with information repetition, Ozkaya and Aslantas propose approaches with triple~\cite{Ozkaya2024} and double~\cite{Ozkaya2025} repetition of recovery information in two separate studies. The first method uses $5\times 5$ blocks and generates 12 authentication bits via SVD, enabling recovery of up to 75\% of the image. The second uses $2\times 2$ blocks, extracting 7 recovery bits from the MSBs of the LL1 coefficient and 5 authentication bits from the first bits of LL1 and the block number. By embedding two copies of recovery information in two partner blocks, this method enables recovery of up to 62.5\% of the tampered image. For real-time applications, Sisaudia and Vishwakarma~\cite{Sisaudia2024} divide the image into 4 sections and generate a 6-bit LBP-based watermark, achieving very fast extraction times (0.47 seconds).

In the domain of chaos-based and logistic map methods, the variety of approaches is notable. Sahu et al.~\cite{Sahu2023} introduced two distinct methods in one study: the first is irreversible and, despite very high visual quality (PSNR $\approx$ 51 dB), does not allow recovery of the original image; in contrast, their second method is reversible, using two mirror images for embedding and enabling full image recovery. In a similar but Walsh-Hadamard-based approach, Adi et al.~\cite{Adi2025} constructed a new adaptive matrix; however, extracting chaos parameters from the image itself makes this method vulnerable to complete forgery. Additionally, Ibrahim et al.~\cite{Ibrahim2025} used Canny edge detection to classify blocks into smooth regions (with $4\times 4$ blocks) and edge/dense regions (with $2\times 2$ blocks), generating watermarks from DWT and logistic maps to achieve a better balance between accuracy and visual quality.

Finally, deep learning-based approaches have also gained attention in this period. Zhang et al.~\cite{Zhang2025} combine active embedding and passive extraction using a degradation-aware network to generate a tamper mask; however, the black-box nature of this method raises concerns about its admissibility in legal and medical applications. Palani and Loganathan~\cite{Palani2024} propose a semi-blind approach using a CoAtNet neural network for feature watermark generation and a turtle shell matrix for embedding. With two-level embedding (feature watermark in the RONI region using TSDH and authentication watermark with DWT-SVD in the 2 LSBs of the U matrix), this method achieves 99.51\% detection accuracy.

\subsection{Categorization and Analysis of Methods}

\subsubsection{Tamper Localization Capability}
According to the survey by Sharma~\cite{Sharma2024}, localization capability is one of the fundamental distinguishing criteria of methods. Methods without localization~\cite{Bouarroudj2024a,Bouarroudj2024b,Akhtarkavan2023}, unless otherwise justified, offer no advantage over strong hash-based methods.

\subsubsection{Presence of a Security Key}
A security key is essential to prevent complete forgery. Methods~\cite{Bouarroudj2024a,Ozkaya2024,Kosuru2023,Ozkaya2025,Adi2025,Sisaudia2024,Akhtarkavan2023} lack a security key and are therefore vulnerable to complete forgery. It should be noted that in~\cite{Bouarroudj2024a}, the authors have used the maximum vector of each block as a key. However, this vector is essentially image data rather than a genuine cryptographic key, and thus it is reproducible. A similar issue exists in~\cite{Ozkaya2024}, where the introduced key is not a true cryptographic key but merely a position-based lookup table.~\cite{Adi2025} also has no independent security key, as its chaotic parameters (P1, P2) are extracted from the image itself and are not a real key.

\subsubsection{Block Size and Its Impact on Accuracy and Security}
According to~\cite{Sharma2024}, the trade-off between block size and localization accuracy is one of the fundamental challenges. Larger blocks reduce localization accuracy, while smaller blocks increase the possibility of targeted attacks. As will be shown in Section 5.13, experimental results of the proposed method on $4\times 4$ block partitioning reveal approximately 12\% duplicate watermarks, while no duplicate watermarks were observed in $6\times 6$ block partitioning.

\subsubsection{Padding Problem and Arbitrary Dimensions}
Most methods assume that the image is divisible by the block size. Some methods, such as~\cite{Akhtarkavan2023}, use edge cropping, while others, like~\cite{Azizoglu2023} and~\cite{Adi2025}, assume compatible dimensions. This assumption limits the scope of application. Any form of padding or cropping violates the principle of full fragility.

\subsubsection{Watermark Generation and Content Dependency}
According to~\cite{Sharma2024}, content dependency is essential for resistance to block-based attacks. Transform-domain methods such as~\cite{Bouarroudj2024a,Bouarroudj2025,Bouarroudj2024b}, which generate watermarks solely from low-frequency DCT coefficients, are vulnerable.

\subsubsection{Resistance to Block-Based Attacks (Collage, Copy-Paste, Copy-Move, VQ)}
According to~\cite{Sharma2024}, for security against block-based attacks, the watermark (authentication bits) of each block must depend on the block's position. Papers lacking position dependency:~\cite{Ibrahim2025,Bouarroudj2024a,Azizoglu2023,Kosuru2023,Sahu2023,Bouarroudj2025,Cedillo2026,Bouarroudj2024b,Sisaudia2024,Wu2026,Akhtarkavan2023}. Papers that claim VQ resistance without substantiation:~\cite{Bouarroudj2024a,Bouarroudj2025,Bouarroudj2024b,Akhtarkavan2023}. Papers confusing VQ with compression:~\cite{Wu2026}. None of the reviewed papers fully demonstrated genuine resistance to VQ. Only~\cite{Palani2024} and~\cite{Zhang2025} partially addressed this issue.

\subsubsection{Fragility Against JPEG Compression}
An ideal fragile watermark should be sensitive to any content modification, including compression, and accurately localize tampered regions. According to Liu et al.~\cite{Liu2025}, repeated or double JPEG compression often indicates tampering and can serve as a forensic trace. Among the reviewed papers,~\cite{Ozkaya2024,Azizoglu2023,Bouarroudj2025,Cedillo2026,Zhang2025,Bouarroudj2024b,Wu2026} addressed this attack, while~\cite{Bouarroudj2024a} and~\cite{Akhtarkavan2023} did not.~\cite{Cedillo2026} and~\cite{Zhang2025} also consider JPEG resistance as an advantage. Given that the method in~\cite{Cedillo2026} explicitly identifies itself as semi-fragile, and the method in~\cite{Zhang2025}—despite its claim of being versatile—practically demonstrates robustness against JPEG compression, both approaches should be classified as semi-fragile.

\subsubsection{Bit-Level Sensitivity}
None of the reviewed papers fully investigated bit-level sensitivity. Transform-domain methods, due to specific sub-band selection, and machine learning-based methods, due to preprocessing, effectively lack this capability. Only~\cite{Zhang2025} indirectly addressed this topic.

\subsubsection{Transform Domain Issues (Rounding Error)}
A fundamental challenge in transform-domain methods is the rounding error. According to Frank Shih's book~\cite{Shih2017}, inverse transform outputs always produce floating-point numbers, and storage in standard formats requires rounding. This error can cause synchronization issues during extraction and may lead to extraction failure even without an attack.

Rounding error occurs at the moment of storage, at the end of the embedding phase, where floating-point values are rounded to integers. Until the image is stored, the embedding process is incomplete, and evaluation without this step merely represents ``laboratory results.'' Therefore, the results of papers that do not address rounding errors are questionable due to incomplete embedding (image storage) and the impossibility of valid extraction. Claims of high PSNR for watermarked images or infinite PSNR for recovered images in these schemes are artifacts of laboratory conditions. For example, PSNR before storage may be very high, but after rounding, it drops to the 51–54 dB range because rounding essentially behaves like a spatial-domain method embedding into one LSB. Furthermore, rounding creates ``synchronization error'' during extraction: the received matrix differs from the original, and the algorithm lacks a common starting point. Consequently, even without an attack, extraction fails. This problem is more acute in reversible watermarking. Optimization-based solutions~\cite{Shih2017} increase execution time. Additionally, rounding error in fragile watermarking does not guarantee single-bit sensitivity and increases forgery risk because rounding is a many-to-one function, allowing an attacker to make changes that do not affect detection.

Among the reviewed papers,~\cite{Bouarroudj2024a,Li2024,Ozkaya2024,Singh2023a,Palani2024,Bouarroudj2025,Cedillo2026,Zhang2025,Ozkaya2025,Bouarroudj2024b} are susceptible to this problem but offer no solution. Paper~\cite{Azizoglu2023} addressed this issue using a DE (Differential Evolution) optimization algorithm to find the best translation matrix for rounding floating-point coefficients to integers with minimal quality loss. However, this approach increases computational cost and requires sending the translation matrix to the receiver. Papers~\cite{Singh2023b} and~\cite{Akhtarkavan2023}, despite being categorized as transform-domain methods, are immune to this problem. They use Integer Wavelet Transform (IWT, also called IIDWT in~\cite{Akhtarkavan2023}), which produces integer outputs and therefore does not suffer from rounding errors.

\subsubsection{Deep Learning Methods: Challenges}
Deep learning-based methods~\cite{Palani2024,Cedillo2026,Zhang2025} face challenges such as black-box nature, lack of legal admissibility, data dependency, and high computational cost~\cite{Hosny2024}. Moreover, machine learning-based methods typically require extensive preprocessing of input images~\cite{Catarino2025}, and this characteristic—which is a critical step for deep learning model performance—fundamentally contradicts the principle of maximum fragility in watermarking, which relies on simplicity and minimal alteration of the original image.

\subsubsection{Bit-Depth and Color Spaces}
Most methods are tested only on 8-bit grayscale images. Papers~\cite{Bouarroudj2024a,Bouarroudj2025,Bouarroudj2024b,Akhtarkavan2023} are designed for high bit-depths, but~\cite{Bouarroudj2024a,Bouarroudj2025,Bouarroudj2024b} lose the ability to detect subtle tampering due to lossy normalization, while~\cite{Akhtarkavan2023} lacks localization capability.

\subsection{Comparison with 7 Selected Papers}
In this subsection, we analyze and compare 7 recent fragile and semi-fragile methods.

In terms of content dependency, six papers~\cite{Adi2025,Wu2026,Sisaudia2024,Bouarroudj2025,Bouarroudj2024a} and partially~\cite{Bouarroudj2024b} use block content, while~\cite{Akhtarkavan2023} uses a fixed watermark. However, against block-based attacks such as VQ, three papers~\cite{Bouarroudj2024b,Bouarroudj2025,Bouarroudj2024a} claim resistance without substantiation. The other four papers~\cite{Adi2025,Wu2026,Sisaudia2024,Akhtarkavan2023} are vulnerable for various reasons. None of these seven papers have proven genuine resistance to VQ. In contrast, our proposed method, with three security layers and empirical verification of watermark uniqueness, guarantees genuine resistance to block-based attacks.

From the perspective of detection metrics, none of the seven papers achieved zero FPR across all attacks. Paper~\cite{Adi2025}, with FPR ranging from 0.0229 to 0.179, and paper~\cite{Bouarroudj2025}, with FPR between 0.001 and 0.08\%, achieved the best performance. Only~\cite{Bouarroudj2025} claimed zero FNR. In terms of bit-level sensitivity, none of the seven papers possess this capability. Paper~\cite{Wu2026}, due to thresholding, ignores changes up to 28 units, and paper~\cite{Sisaudia2024}, due to mean-based LBP, effectively lacks bit-level sensitivity. Regarding JPEG, only~\cite{Bouarroudj2024b} and~\cite{Bouarroudj2024a} thoroughly tested this attack, while~\cite{Bouarroudj2025} performed qualitative analysis. The other four papers~\cite{Adi2025,Wu2026,Sisaudia2024,Akhtarkavan2023} provided no JPEG tests. Table I summarizes the comparison of these seven selected methods across the key criteria discussed above.

\begin{table*}[!t]
\centering
\footnotesize
\caption{Comparison of Seven Selected Fragile Watermarking Schemes}
\label{tab:comparison}
\setlength{\tabcolsep}{4pt}
\begin{tabular}{|l|c|c|p{2.5cm}|p{2.5cm}|c|c|}
\hline
\textbf{Paper} & \textbf{Security Key} & \textbf{Localization} & \textbf{Block Size} & \textbf{Color Space \& Depth} & \textbf{Rounding} & \textbf{Padding} \\
\hline
Bouarroudj 2024 (High Cap.)~\cite{Bouarroudj2024b} & \checkmark & \textbf{X} & $1\times 1$ (Fixed) & Grayscale (8,12,16) & \textbf{X} (DFT) & \textbf{X} \\
\hline
Adi 2025~\cite{Adi2025} & \textbf{X} & \checkmark & Variable (Power-of-2) & Grayscale (8) & \checkmark (Spatial) & \textbf{X} \\
\hline
Wu 2026~\cite{Wu2026} & \checkmark & \checkmark & $4\times 4$ / $2\times 2$ (Fixed) & Grayscale (8) & \checkmark (Spatial) & \textbf{X} \\
\hline
Sisaudia 2024~\cite{Sisaudia2024} & \textbf{X} & \checkmark & $4\times 4$ (Fixed) & Grayscale (8) & \checkmark (Spatial) & \textbf{X} \\
\hline
Akhtarkavan 2023~\cite{Akhtarkavan2023} & \textbf{X} & \textbf{X} & $10\times 10$ (Fixed) & Grayscale (16) & \checkmark (IIDWT) & \textbf{X} \\
\hline
Bouarroudj 2025~\cite{Bouarroudj2025} & \checkmark & \checkmark & $4\times 4$ (Fixed) & Gray/Color (8,12,16) & \textbf{X} (DFT+DWT) & \textbf{X} \\
\hline
Bouarroudj 2024 (Fib.)~\cite{Bouarroudj2024a} & \textbf{X} & \textbf{X} & $4\times 4$ (Fixed) & Grayscale (8,12,16) & \textbf{X} (DFT) & \textbf{X} \\
\hline
\end{tabular}
\end{table*}


Based on the above table, in terms of security keys, only~\cite{Bouarroudj2024b} and~\cite{Bouarroudj2025} use keys but suffer from rounding errors and lack of padding management. Four papers~\cite{Adi2025,Wu2026,Sisaudia2024,Akhtarkavan2023} lack keys but are immune to rounding errors. In terms of localization,~\cite{Adi2025,Wu2026,Sisaudia2024,Bouarroudj2025} have this capability, while~\cite{Bouarroudj2024b,Akhtarkavan2023,Bouarroudj2024a} only provide global detection. Regarding block size, none allow user selection. In terms of bit-depth support, only~\cite{Bouarroudj2025} and~\cite{Bouarroudj2024b} support high bit-depths, but with rounding errors. Padding is not properly managed in any of them.

\subsection{Conclusion and Research Gaps}
Review of 19 recent papers reveals that existing methods face the following fundamental challenges:
\begin{enumerate}
\item \textbf{Non-zero FPR:} None of the methods achieved absolute zero FPR across all attacks.
\item \textbf{Lack of security keys:} Many methods lack keys and are vulnerable to complete forgery.
\item \textbf{Lack of bit-level sensitivity:} None of the methods have investigated single-bit change sensitivity.
\item \textbf{Padding problem:} Most methods depend on fixed and divisible dimensions.
\item \textbf{Fixed block size:} Users cannot choose the block size according to their needs.
\item \textbf{Rounding errors:} Transform-domain methods without proper solutions are impractical.
\item \textbf{Limited bit-depth and color space support:} Most methods are designed only for 8-bit grayscale images.
\end{enumerate}

None of the seven selected papers address all key criteria. Our proposed method, by providing solutions for each of these challenges (security key, precise localization, single-bit sensitivity, arbitrary dimensions, high bit-depth support, user-selectable block size, and full fragility against JPEG), takes a step beyond the current state of the art.


\section{Proposed Fragile Watermarking Scheme}

\subsection{Overview}
The proposed fragile watermarking scheme employs the Triangular Content-Aware Permutation (TCA) as its core scrambling engine. Unlike conventional fragile watermarking methods that rely on fixed mathematical transforms (e.g., parity checks, hash functions, or modular operations), our approach leverages the intrinsic avalanche effect of TCA to generate a content-dependent and highly sensitive watermark for each image block. A single-bit change in the input alters the Delaunay triangulation and, on average, changes more than 50\% of the output bits. The proposed method consists of two main phases: (1) watermark embedding and (2) watermark extraction and tamper localization. Detailed descriptions of these phases are provided in Figs.~1 and 2 and Algorithms~1 and 2, respectively.

\subsection{Preliminaries and Assumptions}
Let $I$ be the original grayscale image of size $M\times N$ pixels. The following parameters are pre-agreed between the sender and receiver over a secure channel:
\begin{itemize}
\item $M, N$: Original image dimensions
\item $m, n$: Block size (typically $m=n=6$, with $m,n\geq 6$)
\item $seed$: Secret seed for pseudo-random noise generation
\item $K$: Optional secondary key for CTR-style multiplication
\item $T$: Number of TCA iterations (default $T=1$)
\end{itemize}
The receiver knows all these parameters a priori.

\subsection{Preprocessing: 7-Bit Image and Noise Masking}
\textbf{Step 1 -- Build 7-bit image:} The least significant bit (LSB) of every pixel is zeroed out using a bitwise AND operation with mask 254:
\begin{equation}
I_7 = I \,\&\, 254
\end{equation}
This reserves the LSB space for watermark embedding while preserving the seven most significant bits (MSBs) that carry the essential visual information.

\textbf{Step 2 -- Generate pseudo-random noise:} Using the secret seed and the Mersenne Twister generator, a noise image $N$ of size $M\times N$ is generated:
\begin{equation}
N = \text{randi}([0, 255], M, N, seed)
\end{equation}

\textbf{Step 3 -- Noise masking via XOR:} The 7-bit image is masked by XORing it with the noise image:
\begin{equation}
I_{\text{mask}} = I_7 \oplus N
\end{equation}
This operation has three critical effects: (a) it eliminates all spatial correlations between blocks; (b) it prevents an attacker from reconstructing the TCA input even if the original image is known; and (c) it makes the watermark dependent on the spatial location of each block through the position-dependent noise pattern.

\subsection{Block Partitioning with Remainder Merging}
Unlike conventional block-based methods that require the image dimensions to be multiples of the block size (or rely on padding), our scheme employs a remainder-merging strategy that supports arbitrary image dimensions.

The number of blocks in the vertical and horizontal directions is calculated as:
\begin{equation}
B_v = \max(1, \lfloor M/m \rfloor), \qquad B_h = \max(1, \lfloor N/n \rfloor)
\end{equation}

For blocks not in the last row or last column, the size is exactly $m\times n$. For blocks in the last row (or last column), the remaining rows (or columns) are merged into that block, making it larger. The bottom-right block absorbs all remaining pixels in both directions. This strategy preserves image integrity, introduces no artificial padding values, and ensures that every block is at least $m\times n$ (with $m,n\geq 6$), which is necessary for TCA scrambling.

\subsection{CTR-Style Block-Dependent Key Generation}
To prevent block relocation attacks (e.g., cut-and-paste or collage attacks), each block is processed with a unique multiplier derived from a counter-mode (CTR) approach. For the block with linear index $idx$ (scanning row-wise from 1 to $B_v \times B_h$), the multiplier is:
\begin{equation}
\lambda_{idx} = K + (idx - 1)
\end{equation}

This multiplier is applied element-wise to the corresponding block $B_{\text{mask}}$ from $I_{\text{mask}}$:
\begin{equation}
B_{\text{mult}} = ( \text{double}(B_{\text{mask}}) \times \lambda_{idx} ) \bmod 256
\end{equation}

The result is then scaled back to the $[0, 255]$ range. This operation ensures that even identical pixel blocks at different spatial positions receive completely different numerical values, and thus generate different watermarks.

\subsection{Bit-Plane Extraction and Bit-Matrix Formation}
The multiplied block $B_{\text{mult}}$ of size $b_h\times b_w$ (where $b_h$ and $b_w$ are the actual block dimensions after merging) is converted into a binary bit-matrix. First, all eight bit-planes are extracted:
\begin{equation}
c_1 = \text{bitget}(B_{\text{mult}}, 1),\quad \ldots, \quad c_8 = \text{bitget}(B_{\text{mult}}, 8)
\end{equation}

These eight bit-planes are then concatenated to form a single binary matrix:
\begin{equation}
B_{\text{bits}} = [c_1, c_2, c_3, c_4 ; c_5, c_6, c_7, c_8]
\end{equation}

The resulting matrix has dimensions $(2\times b_h) \times (4\times b_w)$. For a standard $6\times 6$ block, this yields a $12\times 24$ binary matrix containing exactly 288 bits.

\subsection{TCA-Based Bit Scrambling}
The binary matrix $B_{\text{bits}}$ is fed into the TCA (Triangular Content-Aware Permutation) algorithm. TCA is a pure permutation cipher that rearranges the positions of bits without modifying their values. Its operation consists of three stages:
\begin{enumerate}
\item \textbf{Edge detection:} The Canny edge detector is applied to the binary matrix to identify edge points (transitions between 0 and 1).
\item \textbf{Delaunay triangulation:} The detected edge points, together with the four corner points, form a Delaunay triangulation.
\item \textbf{Bit rearrangement:} Bits are extracted by traversing the triangles and then reshaped back into a matrix of the original dimensions.
\end{enumerate}

The scrambling is applied for a pre-agreed number of iterations $T$:
\begin{equation}
B_{\text{scram}} = \text{TCA}^{(T)}(B_{\text{bits}})
\end{equation}

The key property of TCA for our scheme is its intrinsic avalanche effect: a single-bit change in the input binary matrix causes a completely different Delaunay triangulation, which in turn alters the permutation order and changes a significant fraction of the output bits.

\subsection{Watermark Selection and LSB Embedding}
From the scrambled binary matrix $B_{\text{scram}}$ of size $(2\times b_h) \times (4\times b_w)$, we select a central submatrix of size $b_h\times b_w$ as the watermark:
\begin{equation}
W = B_{\text{scram}}(2 : 1+b_h , 2 : 1+b_w)
\end{equation}

The offset of one row and one column from the corners avoids potential boundary artifacts of the Delaunay triangulation. The number of selected bits exactly equals the number of pixels in the original block ($b_h \times b_w$).

Although TCA as a pure permutation in its complete form (without truncation) has no collisions, the final watermark is selected from these bits. For a $6\times 6$ block, selecting 36 bits from the 288-bit output constitutes the block watermark. This truncation is the only potential source of collisions, and experiments show that its probability for targeted forgery is negligible (FNR = 0.27\% only for salt-and-pepper noise).

Furthermore, block size selection directly affects watermark uniqueness. Although the proposed method is theoretically applicable to any block size (even $1\times 1$), for small blocks (especially smaller than $6\times 6$), the output space becomes limited and duplicate watermarks appear. Experimental results show that for $4\times 4$ blocks, more than 12\% of watermarks are duplicates. Therefore, it is recommended to choose a block size of $6\times 6$ or larger, as no duplicate watermarks are observed in this range.

Finally, the watermark is embedded into the LSB of the corresponding block from the original 7-bit image $I_7$ using the bitset operation:
\begin{equation}
B_{\text{wm}} = \text{bitset}(B_7, 1, W)
\end{equation}

The watermarked blocks are then reassembled into the final watermarked image $I_W$ using the inverse of the remainder-merging strategy.

Fig.~1 shows the flowchart of the embedding algorithm. Algorithm~1 presents its step-by-step details.

\begin{figure*}[!t]
\centering
\includegraphics[width=3in]{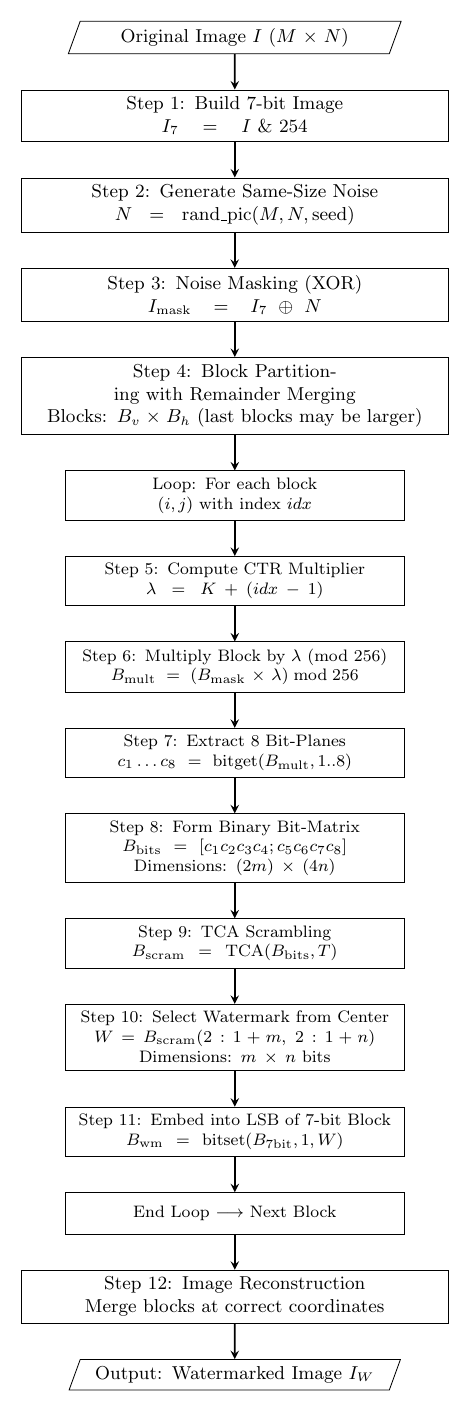}
\caption{Flowchart of the watermark embedding algorithm. The main steps include preprocessing (7-bit image construction, noise generation, and masking), block partitioning, CTR multiplication, bit-plane extraction, TCA scrambling, central watermark selection, and LSB embedding.}
\label{fig:embed}
\end{figure*}

\begin{algorithm}[!t]
\caption{Watermark Embedding}
\label{alg:embed}
\begin{algorithmic}[1]
\Require $I$ (original image), $M,N,m,n,seed,K,T$
\Ensure $I_W$ (watermarked image)

\State $I_7 \gets \text{bitand}(I, 254)$
\State $N \gets \text{rand\_pic\_uint8\_seeded1}(M,N,seed)$
\State $I_{\text{mask}} \gets I_7 \oplus N$

\State $\text{blocks}_{\text{mask}} \gets \text{divideImage\_MergeRemainder}(I_{\text{mask}}, m, n)$
\State $\text{blocks}_7 \gets \text{divideImage\_MergeRemainder}(I_7, m, n)$

\State $\text{blocks}_{\text{wm}} \gets \text{cell array of same size}$

\For{each block with index $idx$ (row-major)}
    \State $B_{\text{mask}} \gets \text{blocks}_{\text{mask}}[idx]$
    \State $B_7 \gets \text{blocks}_7[idx]$
    \State $[b_h, b_w] \gets \text{size}(B_{\text{mask}})$
    \State $\lambda \gets K + (idx - 1)$
    \State $B_{\text{mult}} \gets (B_{\text{mask}} \times \lambda) \bmod 256$
    
    \State $c_1 \ldots c_8 \gets \text{bitget}(B_{\text{mult}}, 1 \ldots 8)$
    \State $B_{\text{bits}} \gets [c_1, c_2, c_3, c_4; c_5, c_6, c_7, c_8]$
    
    \State $B_{\text{scram}} \gets \text{TCA}^{(T)}(B_{\text{bits}})$
    
    \State $W \gets B_{\text{scram}}(2:1+b_h,\ 2:1+b_w)$
    \State $B_{\text{wm}} \gets \text{bitset}(B_7, 1, W)$
    \State $\text{blocks}_{\text{wm}}[idx] \gets B_{\text{wm}}$
\EndFor

\State $I_W \gets \text{mergeImage\_MergeRemainder}(\text{blocks}_{\text{wm}}, [M, N])$
\State \Return $I_W$
\end{algorithmic}
\end{algorithm}

\subsection{Watermark Extraction and Tamper Localization}
The extraction phase is performed by the receiver without requiring the original image. The receiver knows all security parameters ($M, N, m, n, seed, K, T$) and receives the possibly tampered image $I_{\text{rec}}$.

\textbf{Step 1 -- Extract existing watermark:} The LSB of each pixel is extracted to obtain the embedded watermark:
\begin{equation}
W_{\text{ext}} = \text{bitget}(I_{\text{rec}}, 1)
\end{equation}

\textbf{Step 2 -- Recompute watermark from 7 MSBs:} The receiver follows the same steps as the embedding phase, but using the 7 MSBs of $I_{\text{rec}}$ (after zeroing its LSB) instead of the original image. This produces a recalculated watermark $W_{\text{calc}}$ for each block.

\textbf{Step 3 -- Compare:} For each block $(i, j)$ with size $b_h\times b_w$:
\begin{itemize}
\item If $W_{\text{ext}} = W_{\text{calc}}$, the block is declared authentic.
\item Otherwise, the block is declared tampered.
\end{itemize}

\textbf{Step 4 -- Tamper localization:} When a mismatch is detected, the spatial region of that block is marked as tampered. A binary tamper map $T_{\text{map}}$ of size $M\times N$ is constructed, where $T_{\text{map}}(r,c)=0$ indicates a tampered pixel and $1$ indicates an authentic pixel. Localization granularity equals the block size ($m\times n$ pixels).

Extraction is single-phase: detection and localization occur simultaneously. Since each block carries its own independent watermark, the scheme can precisely pinpoint which blocks have been tampered.

Fig.~2 shows the flowchart of the extraction and tamper localization algorithm. Algorithm~2 presents its step-by-step details.

\begin{figure*}[!t]
\centering
\includegraphics[width=6in]{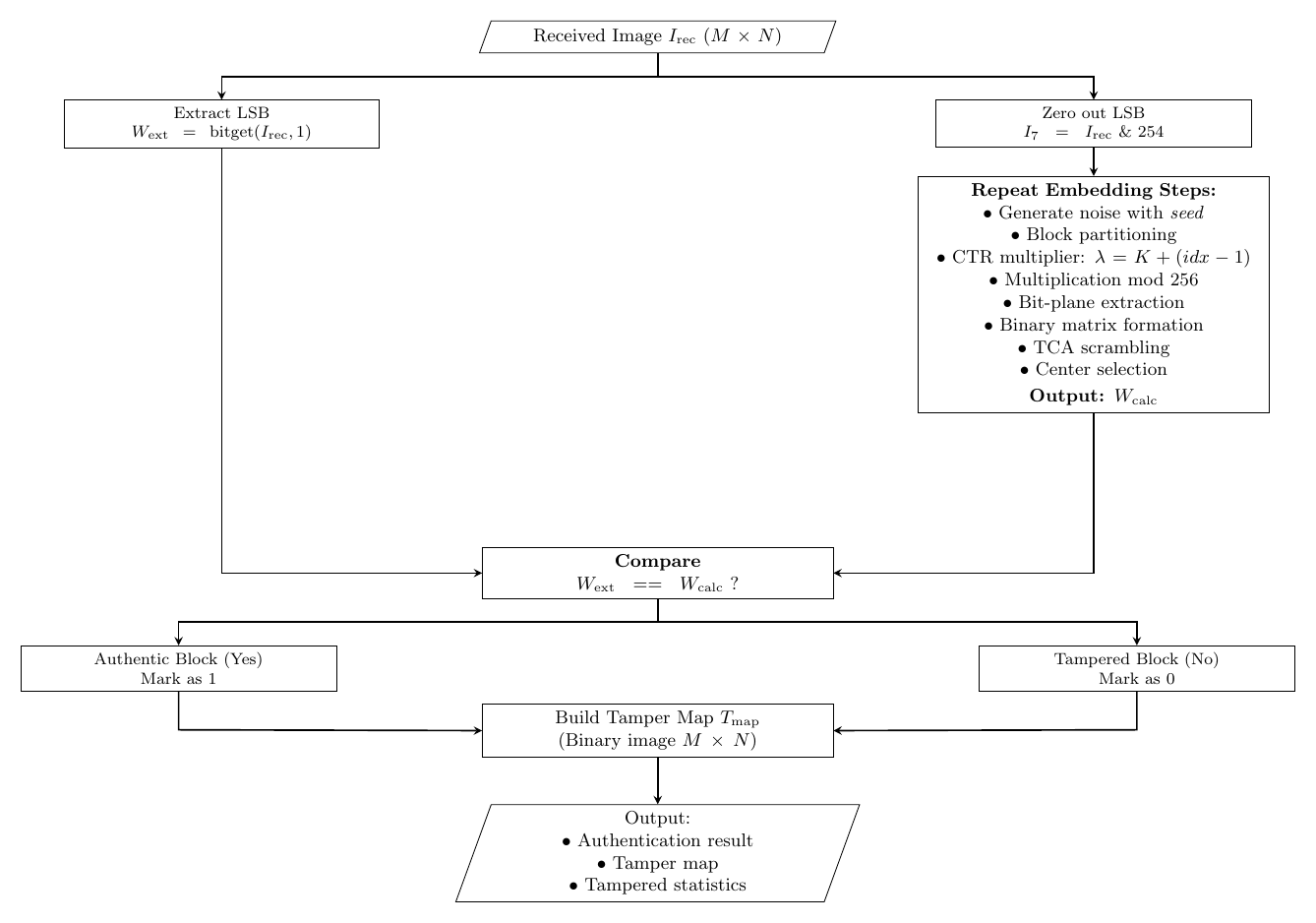}
\caption{Flowchart of the watermark extraction and tamper localization algorithm. The steps include extracting the existing watermark, recomputing the watermark from the 7 MSBs, comparing, and generating the binary tamper map.}
\label{fig:extract}
\end{figure*}

\begin{algorithm}[!t]
\caption{Watermark Extraction and Tamper Localization}
\label{alg:extract}
\begin{algorithmic}[1]
\Require $I_{\text{rec}}$ (received image), $M,N,m,n,seed,K,T$
\Ensure $T_{\text{map}}$ (tamper map, 1=authentic, 0=tampered), $auth$ (authentication result)

\State $W_{\text{ext}} \gets \text{bitget}(I_{\text{rec}}, 1)$
\State $I_7 \gets \text{bitand}(I_{\text{rec}}, 254)$

\State $N \gets \text{rand\_pic\_uint8\_seeded1}(M, N, seed)$
\State $I_{\text{mask}} \gets I_7 \oplus N$

\State $\text{blocks}_{\text{ext}} \gets \text{divideImage\_MergeRemainder}(W_{\text{ext}}, m, n)$
\State $\text{blocks}_{\text{mask}} \gets \text{divideImage\_MergeRemainder}(I_{\text{mask}}, m, n)$

\State $T_{\text{map}} \gets \text{ones}(M, N)$
\State $auth \gets \text{true}$

\For{each block with index $idx$ (row-major order)}
    \State $B_{\text{mask}} \gets \text{blocks}_{\text{mask}}[idx]$
    \State $[b_h, b_w] \gets \text{size}(B_{\text{mask}})$
    \State $\lambda \gets K + (idx - 1)$
    \State $B_{\text{mult}} \gets \text{mod}(B_{\text{mask}} \times \lambda, 256)$
    
    \State Extract $c_1 \ldots c_8 \gets \text{bitget}(B_{\text{mult}}, 1 \ldots 8)$
    \State $B_{\text{bits}} \gets [c_1, c_2, c_3, c_4; c_5, c_6, c_7, c_8]$
    \State $B_{\text{scram}} \gets \text{TCA}(B_{\text{bits}}, T)$
    \State $W_{\text{calc}} \gets B_{\text{scram}}(2:1+b_h, 2:1+b_w)$
    
    \If{$W_{\text{calc}} \neq \text{blocks}_{\text{ext}}[idx]$}
        \State $auth \gets \text{false}$
        \State Mark all pixels of this block as 0 in $T_{\text{map}}$
    \EndIf
\EndFor

\State \Return $T_{\text{map}}$, $auth$
\end{algorithmic}
\end{algorithm}

\subsection{Security Discussion}
The proposed scheme achieves security through multiple layers of dependency, as summarized in Table I.

\begin{table*}[!t]
\centering
\footnotesize
\caption{Security Layers of the Proposed Scheme}
\label{tab:security}
\setlength{\tabcolsep}{3pt}
\begin{tabular}{|p{2.8cm}|p{4.5cm}|p{4.5cm}|}
\hline
\textbf{Layer} & \textbf{Mechanism} & \textbf{Purpose} \\
\hline
Noise masking & XOR with seed-driven noise & Spatial location binding \\
\hline
CTR-style keying & Multiplication by $K+(idx-1)$ & Block order binding \\
\hline
TCA scrambling & Content-dependent permutation & Avalanche effect + non-analytic behavior \\
\hline
Center selection & Fixed offset from corners & Avoid boundary artifacts \\
\hline
\end{tabular}
\end{table*}

If an attacker compromises one key layer (e.g., the noise seed), the other layer (e.g., $K$) still blocks block relocation and VQ attacks. Moreover, the content-dependent nature of TCA ensures that the permutation pattern for one block reveals no information about the pattern for any other block, fundamentally eliminating the vulnerability of dimension-based scramblers (such as Arnold, Zigzag, or Spiral transforms) to known-plaintext attacks.

\subsection{Extension to Color Images (RGB)}

The proposed method, described so far for grayscale images, can be easily extended to RGB color images without any change to the core algorithm. The main idea is to convert the three-dimensional structure of the color image into a two-dimensional structure while preserving the spatial dependency between color channels.

\subsubsection{Vertical Sandwiching Transformation}
\textbf{Definition 1 (Vertical Sandwiching).} Let $I_{\text{RGB}}$ be a color image with dimensions $H\times W\times 3$. The sandwich function $S: \mathbb{R}^{H\times W\times 3} \rightarrow \mathbb{R}^{3H\times W}$ is defined as:
\begin{equation}
S(I_{\text{RGB}}) = [I_R; I_G; I_B]
\end{equation}
where $I_R$, $I_G$, $I_B$ are the red, green, and blue channels, respectively. Thus, each $m\times n$ block in the original image is transformed into a $(3m)\times n$ block in the sandwiched image, containing information from all three color channels for that spatial region. Fig.~3 illustrates the vertical sandwiching process.

\begin{figure}[!t]
\centering
\includegraphics[width=3.7in]{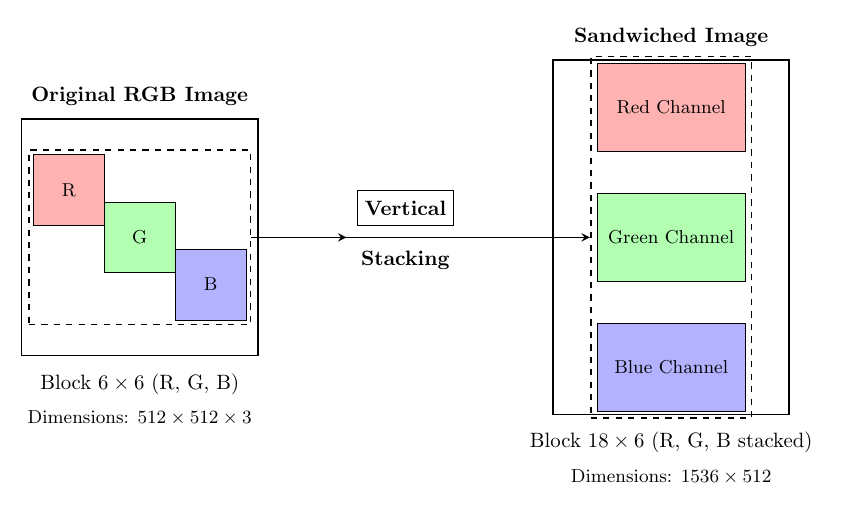}
\caption{Vertical sandwich transformation for color images. The three R, G, B channels are stacked vertically to form a grayscale-like image of size $3H\times W$. Each $m\times n$ block from the original image is transformed into a $(3m)\times n$ block in the sandwiched image.}
\label{fig:sandwich}
\end{figure}

\subsubsection{Embedding and Extraction Algorithm for Color Images}
After applying function $S$:
\begin{enumerate}
\item The embedding and extraction algorithms run exactly as in the grayscale case on the resulting image with dimensions $3H\times W$. All functions including divideImage\_MergeRemainder, TCA scrambling, and bitwise operations work without any modification.
\item The block size changes from $m\times n$ to $(3m)\times n$. However, the number of blocks remains unchanged (equal to $\lceil H/m \rceil \times \lceil W/n \rceil$).
\item Security parameters (secret seed, key $K$, and number of TCA iterations) are the same for all three channels.
\item During reconstruction, the inverse of function $S$ is applied to recover the color watermarked image.
\end{enumerate}

\subsubsection{Practical Implementation}
In practical implementation, two helper functions are defined:
\begin{itemize}
\item \texttt{color\_to\_block\_sandwich}: converts a color image of size $H\times W\times 3$ into a single image of size $(3H)\times W$.
\item \texttt{block\_sandwich\_to\_color}: converts a single image of size $(3H)\times W$ back into a color image of size $H\times W\times 3$.
\end{itemize}

After this transformation, the embedding and extraction codes can be used without any changes.

Table II summarizes the advantages of the proposed approach for color images.

\begin{table*}[!t]
\centering
\footnotesize
\caption{Advantages of the Proposed Approach for Color Images}
\label{tab:color}
\setlength{\tabcolsep}{4pt}
\begin{tabular}{|p{3.5cm}|p{11cm}|}
\hline
\textbf{Advantage} & \textbf{Explanation} \\
\hline
Zero change to core algorithm & All functions including block partitioning, TCA scrambling, and bitwise operations work without any modification. \\
\hline
Preserves inter-channel dependency & Unlike methods that process each color channel independently~\cite{Ibrahim2025,Cedillo2026,Palani2024,Bouarroudj2025,Wu2026}, the proposed approach maintains dependency between channels. A manipulation in one channel (e.g., only the red channel) changes the watermark of the entire block. \\
\hline
Uniform security parameters & The secret seed, key $K$, and number of TCA iterations are the same for all three channels. \\
\hline
Favorable time efficiency & The number of blocks remains constant and only the block height increases. Consequently, execution time increases by about 1.7 times (not 3 times). In contrast, if the three color channels were processed independently, the number of blocks would triple and execution time would approximately triple as well. \\
\hline
\end{tabular}
\end{table*}

To verify the effectiveness of the method on color images, a set of 10 standard color images from MATLAB R2021b default images (including autumn, trees, fabric, flamingos, llama, peacock, hallway, indiancorn, kids, monkey) were resized to $512\times 512$ pixels. Experimental results are presented in Section 5.10 and show that the proposed method performs similarly to (and in some cases better than) the grayscale case.

\subsection{Extension to High Bit-Depth Images}
High bit-depth images (12 or 16 bits) are widely used in sensitive domains such as medical imaging (CT scan, MRI, X-ray) and remote sensing. In these images, each pixel stores more precise information than ordinary 8-bit images, and preserving all details is critical for accurate diagnosis. However, many existing fragile watermarking methods~\cite{Bouarroudj2025} normalize high bit-depth images to 8 bits before processing. This operation discards lower bits and creates many-to-one mappings, allowing an attacker to make subtle but important changes to the original image without the watermark being detected.

Our proposed method is inherently free from this problem and can work directly on high bit-depth images with three simple modifications compared to the 8-bit version:

\textbf{Step 1 -- Adjust modulo range:} In the 8-bit version, CTR multiplication uses modulo 256. For bit-depth $d$, this operation changes to modulo $2^d$:
\begin{equation}
B_{\text{mult}} = ( \text{double}(B_{\text{mask}}) \times \lambda_{idx} ) \bmod 2^d
\end{equation}
The result is then scaled back to the $[0, 2^d-1]$ range, preserving the full dynamic range of the original image.

\textbf{Step 2 -- Bit-matrix structure:} For 8-bit images, the bit-matrix is formed as $[c_1 \ldots c_4; c_5 \ldots c_8]$. For higher bit-depths, all $d$ bit-planes are extracted and stacked vertically:
\begin{itemize}
\item For 12-bit: $[c_1 \ldots c_4; c_5 \ldots c_8; c_9 \ldots c_{12}]$
\item For 16-bit: $[c_1 \ldots c_4; c_5 \ldots c_8; c_9 \ldots c_{12}; c_{13} \ldots c_{16}]$
\end{itemize}
The resulting matrix dimensions become $(\lceil d/4 \rceil \times b_h) \times (4 \times b_w)$. For a $6\times 6$ block in the 16-bit case, a $24\times 24$ matrix containing 576 bits is obtained.

\textbf{Step 3 -- Bit-depth aware noise generation:} The pseudo-random noise image is generated in the range $[0, 2^d-1]$ using the same secret seed:
\begin{equation}
N = \text{rand\_pic\_uint16\_seeded}(M, N, seed, d)
\end{equation}

All other steps of the algorithm -- including noise masking (XOR), remainder-merging block partitioning, TCA scrambling, center watermark selection, and extraction -- remain unchanged. This demonstrates the inherent scalability of the TCA-based approach to arbitrary bit-depths without requiring lossy normalization or conversion.

\textbf{Note on advantages over existing methods:} Unlike the method proposed by Bouarroudj et al.~\cite{Bouarroudj2025}, which normalizes 16-bit images to 8-bit before watermark generation and embedding (and thus, at best, operates on an 8-bit image), our method works directly on the original bit-depth without any bit-depth reduction. Consequently, every single-bit change in any bit-plane (from bit 0 to bit 15) is detected with perfect accuracy. This makes the proposed scheme significantly more suitable for high bit-depth medical images, where preserving the full dynamic range of pixel values is clinically critical.

\subsection{Note on Parallelization Capability}

It is worth noting that the proposed method inherently supports full block-level parallelization. Since the global noise masking (which provides both location dependency and inter-block dependency) is applied before the main block processing loop, every block can be processed independently and concurrently without compromising security. This stands in sharp contrast to hierarchical or chain-based methods, where the watermark of each block explicitly depends on neighboring blocks, forcing sequential processing. The combination of the intrinsic avalanche effect of TCA (which ensures that any change in block content produces a completely different watermark) and the XOR-based noise masking (which ensures that blocks from different images or different locations are not reusable) enables this independence without sacrificing security. This feature alone constitutes an independent innovation, as it simultaneously achieves the seemingly conflicting objectives of strong inter-block security dependency and complete computational independence for parallel execution. Experimental results in Section 5.6 demonstrate a 3.9$\times$ speedup (from 6.36 s to 1.61 s) using a straightforward parallel loop implementation.
\section{Experimental Results}

\subsection{Experimental Environment and Dataset}
The experiments were conducted on a system with Windows 11 Pro, an Intel Core i5-1335U processor (13th generation, 1.30 GHz, 10 cores, 12 logical threads), 16 GB of RAM, and MATLAB R2021b.

The dataset consists of 50 standard grayscale images of size $512\times 512$. Forty-eight images were selected from the UGR database, and the remaining two were obtained from USC-SIPI (converted to grayscale) and the default MATLAB image (converted to grayscale and resized to $512\times 512$).

The fixed experimental parameters were: $M=N=512$, $m=n=6$, $seed = 123456$, $K=21.2$, and number of TCA iterations $T=1$. All mentioned parameters (image dimensions, block size, secret seed, key $K$, and number of iterations) are adjustable; the above values were chosen solely for these experiments.

\subsection{Evaluation Metrics}
To evaluate detection performance, standard metrics based on TP (tampered pixels correctly detected), TN (authentic pixels correctly identified), FP (authentic pixels falsely flagged as tampered), and FN (tampered pixels falsely flagged as authentic) are used:

\begin{itemize}
\item \textbf{True Positive Rate (TPR) / Sensitivity:} $TPR = TP/(TP+FN)$
\item \textbf{False Positive Rate (FPR):} $FPR = FP/(FP+TN)$
\item \textbf{False Negative Rate (FNR):} $FNR = FN/(TP+FN)$
\item \textbf{Accuracy (ACC):} $ACC = (TP+TN)/(TP+TN+FP+FN)$
\item \textbf{Precision:} $Precision = TP/(TP+FP)$
\item \textbf{F1-Score:} $F1 = (2 \times Precision \times TPR)/(Precision + TPR)$
\end{itemize}

In fragile watermarking, an ideal detector should achieve $TPR=100\%$, $FPR=0\%$, $FNR=0\%$, and $ACC=100\%$, indicating complete detection of tampered pixels and no false alarms for authentic pixels.

\subsection{Attack Implementation}
Watermarked images were subjected to the following attacks. Except for the Photoshop collage attack (on two images) and salt-and-pepper noise (repeated three times per image), all attacks were applied to all 50 images. Detection was performed block-wise by comparing the watermark extracted from the LSB with the watermark recomputed from the 7 MSBs.

\begin{itemize}
\item \textbf{Gaussian noise:} Using \texttt{imnoise}, mean 0, variances $\sigma^2 \in \{0.0001, 0.0005, 0.001, 0.005, 0.01, 0.02, 0.05, 0.1\}$.
\item \textbf{Poisson noise:} Using \texttt{imnoise('poisson')}.
\item \textbf{Speckle noise:} Using \texttt{imnoise('speckle')}, densities $d \in \{0.002, 0.02, 0.2\}$.
\item \textbf{Salt-and-pepper noise:} Using \texttt{imnoise}, densities $d \in \{0.0001, 0.0005, 0.001, 0.005, 0.01, 0.02, 0.05, 0.1\}$. Each density was repeated three times.
\item \textbf{Brightness adjustment:} Using \texttt{imadd}, deltas $\Delta \in \{1, 10, 50, 100, 149, 200, 243, 254, -254\}$.
\item \textbf{Contrast adjustment:} Using \texttt{imadjust}, factors $\gamma \in \{0.1, 0.2, 0.5, 0.8, 0.9\}$.
\item \textbf{Median filtering:} Using \texttt{medfilt2}, $3\times 3$ neighborhood.
\item \textbf{Rotation:} Angles $1^\circ, 90^\circ, 180^\circ$ using \texttt{imrotate} (with 'loose' and 'crop' options) and resizing back to $512\times 512$.
\item \textbf{Ellipse region removal:} A random ellipse was filled with white pixels (255) (random center, axes $a, b$).
\item \textbf{Line addition:} A diagonal line (main diagonal) was set to zero.
\item \textbf{JPEG compression:} Using \texttt{imwrite}, quality factors $Q \in \{5, 10, 20, 30, 40, 50, 60, 70, 80, 90, 95\}$.
\item \textbf{Block mean replacement:} Each block was replaced with a matrix of its mean value.
\item \textbf{Copy-move (intra-image):} A random block was copied and pasted from one location to another.
\item \textbf{Collage (different K):} Two images with different $K$ values (same seed). A random block from the first image replaced the corresponding block in the second.
\item \textbf{Collage (different seed):} Two images with different seeds (same $K$). A random block from the first image replaced the corresponding block in the second.
\item \textbf{Manual Photoshop collage:} An image was manipulated using Adobe Photoshop (copying, moving, removing elements) to appear natural (e.g., a bird from one leg to two). Tested on two images.
\item \textbf{Complete random pixel value change:} A random pixel was changed to a new random value. This tests sensitivity to arbitrary pixel changes.
\item \textbf{Single-bit flipping:} A random pixel and a random bit position (0 to 7) were flipped using \texttt{random\_bit\_change}. This tests bit-level sensitivity, including LSB modifications.
\end{itemize}

\subsection{Quantitative Results and Analysis}
Table III summarizes the detection performance against all 18 attacks. For 17 attacks, the method achieved perfect detection: $TPR=100\%$, $FPR=0\%$, $FNR=0\%$, $ACC=100\%$.

\begin{table*}[!t]
\centering
\footnotesize
\caption{Detection Performance against Different Attacks}
\label{tab:detection}
\setlength{\tabcolsep}{3pt}
\begin{tabular}{|l|c|c|c|c|c|c|}
\hline
\textbf{Attack} & \textbf{TPR (\%)} & \textbf{FPR (\%)} & \textbf{FNR (\%)} & \textbf{ACC (\%)} & \textbf{Precision} & \textbf{F1} \\
\hline
Gaussian Noise & 100 & 0 & 0 & 100 & 1 & 1 \\
\hline
Poisson Noise & 100 & 0 & 0 & 100 & 1 & 1 \\
\hline
Speckle Noise & 100 & 0 & 0 & 100 & 1 & 1 \\
\hline
Brightness Adjustment & 100 & 0 & 0 & 100 & 1 & 1 \\
\hline
Contrast Adjustment & 100 & 0 & 0 & 100 & 1 & 1 \\
\hline
Median Filtering & 100 & 0 & 0 & 100 & 1 & 1 \\
\hline
Rotation & 100 & 0 & 0 & 100 & 1 & 1 \\
\hline
Ellipse Removal & 100 & 0 & 0 & 100 & 1 & 1 \\
\hline
Line Addition & 100 & 0 & 0 & 100 & 1 & 1 \\
\hline
JPEG Compression & 100 & 0 & 0 & 100 & 1 & 1 \\
\hline
Block Mean Replacement & 100 & 0 & 0 & 100 & 1 & 1 \\
\hline
Copy-Move & 100 & 0 & 0 & 100 & 1 & 1 \\
\hline
Collage (Different K) & 100 & 0 & 0 & 100 & 1 & 1 \\
\hline
Collage (Different Seed) & 100 & 0 & 0 & 100 & 1 & 1 \\
\hline
Photoshop Collage & 100 & 0 & 0 & 100 & 1 & 1 \\
\hline
Random Pixel Change & 100 & 0 & 0 & 100 & 1 & 1 \\
\hline
Single-Bit Flipping & 100 & 0 & 0 & 100 & 1 & 1 \\
\hline
Salt-and-Pepper Noise & 99.73 & 0 & 0.27 & 99.97 & 1 & 0.999 \\
\hline
\end{tabular}
\end{table*}

\textbf{Analysis:} Perfect detection for 17 attacks confirms that the method is highly sensitive to any type of manipulation, whether global (brightness, contrast, rotation, JPEG) or local (copy-move, collage, ellipse removal, line addition, mean replacement, pixel changes, single-bit flipping). The content-dependent watermark from TCA scrambling ensures that any change in the 7 MSBs—no matter how small—produces a completely different recomputed watermark that does not match the extracted one.

The only attack with a non-zero FNR was salt-and-pepper noise. This is expected since the noise randomly converts pixels to 0 or 255. FNR occurs only when the noise changes enough MSB bits but coincidentally produces a watermark that matches the extracted one. This probability is very low, as confirmed by $TPR=99.73\%$ and $FNR=0.27\%$. Importantly, $FPR=0\%$ remains, meaning no authentic pixels are falsely flagged as tampered.

\subsection{Security Analysis: Resistance to VQ and Collage Attacks}
A common vulnerability in block-based schemes is susceptibility to VQ and collage attacks, where the attacker builds a codebook of valid watermarks. The proposed method is inherently resistant to such attacks, as demonstrated by theoretical analysis and empirical verification.

\textbf{Empirical Uniqueness Verification:} A pairwise distinctiveness test was performed on all watermarks. For 10 images (approximately 72,820 blocks of size $6\times 6$), all watermarks were generated and every pair was compared. Results: (1) within a single image, no two blocks produced identical watermarks—making copy-move (replacing one block with another in the same image) impossible; (2) across different images, no two blocks produced identical watermarks—making collage (replacing a block from another image) impossible, even with identical pixels.

This result, combined with the following theoretical analysis, proves that an attacker cannot build a codebook. First, unique content-dependent watermarks: For a $6\times 6$ block, the watermark is derived from the center of the $12\times 24$ scrambled bit-matrix (from 288 bits of block content). No two blocks—even with identical pixels—produce the same watermark due to noise masking (XOR with seed) and the CTR multiplier ($\lambda_{idx}=K+(idx-1)$). Second, non-analytic permutation: The TCA pattern is content-dependent and non-analytic. The Delaunay triangulation changes drastically with any modification. The mapping from block content to watermark is not a fixed function that can be precomputed. Without knowledge of the seed and $K$, the attacker cannot generate a valid watermark, making codebook construction computationally infeasible. Thus, VQ, collage, block relocation, copy-paste, and codebook-based forgery attacks are effectively neutralized.

\subsection{Sensitivity to Single-Bit Changes}
To assess maximum sensitivity, a single-bit experiment was designed (since the smallest possible change in a pixel is flipping one bit). A random pixel was selected and exactly one bit (0 to 7) was flipped. Results showed that in all cases, the tampering was detected due to the intrinsic avalanche effect of TCA. The experiment on 50 images resulted in $TPR=100\%$ and $FPR=0\%$, confirming that the method is sensitive to any single-bit change, regardless of the bit position.

\subsection{Computational Cost and Visual Quality}
\textbf{Execution Time:} The average embedding time was 1.61 seconds, and extraction time was 1.63 seconds over 50 images, using parallel processing with 10 workers on an Intel Core i5-1335U processor (10 physical cores, 12 logical threads). The parallel implementation leverages the inherent block-wise independence of the algorithm, achieving a speedup of approximately 3.9$\times$ compared to the single-threaded version (6.36 s vs. 1.61 s). All measurements were performed on a laptop CPU without GPU acceleration. This execution time is not only acceptable for offline fragile watermarking applications, where embedding is performed once and verification is done offline, but also makes the method suitable for near real-time applications where rapid authentication is required.

\textbf{Visual Quality:} The method only modifies the LSB and preserves the 7 MSBs, resulting in minimal visual distortion. The average Peak Signal-to-Noise Ratio (PSNR) between the original and watermarked images was 51.14 dB, and the average Structural Similarity Index (SSIM) was 0.9975, indicating excellent quality and structural preservation. These values align with theoretical expectations for LSB-based embedding schemes.

\subsection{Discussion and Comparative Summary}
The experimental results demonstrate that the proposed method achieves:
\begin{itemize}
\item Perfect detection ($TPR=100\%$, $FPR=0\%$, $FNR=0\%$) for 17 out of 18 attacks, including global manipulations (brightness, contrast, rotation, JPEG), local manipulations (copy-move, collage, ellipse removal, line addition, pixel changes, single-bit flipping), and common operations (median filtering, mean replacement).
\item Highly robust detection for salt-and-pepper noise ($TPR=99.73\%$, $FPR=0\%$), with $FNR=0.27\%$ due to the random nature of the noise.
\item Strong theoretical resistance to VQ and collage attacks through content-dependent watermarks and non-analytic permutation.
\item Sensitivity to single-bit changes, regardless of bit position.
\item Imperceptible embedding ($PSNR=51.14$ dB, $SSIM=0.9975$) and practical computational cost (1.61 s embedding, 1.63 s extraction on a standard laptop).
\end{itemize}

Compared to conventional schemes relying on fixed functions or parity checks, the proposed method offers superior sensitivity while maintaining visual quality. The use of TCA as a pure permutation cipher eliminates the need for separate diffusion stages and achieves avalanche sensitivity through content-dependent scrambling. The results confirm that the method is both highly sensitive to manipulation and robust against advanced attacks such as VQ and collage.

\subsection{Analysis of Negligible FNR: Why It Is Not Exploitable for Forgery}
As observed in Table III, the only attack with a non-zero FNR (0.27\%) was salt-and-pepper noise. In this section, we show that this negligible FNR differs from the high FNR in shortened hash-based methods and is not exploitable for targeted forgery.

\textbf{Why does our method have near-zero FNR?} TCA is a pure permutation with an intrinsic avalanche effect. It is bijective and, unlike hash functions, inherently collision-free. The watermark is selected as 36 bits from the total scrambled output (288 bits for a $6\times 6$ block). This truncation is the only source of collisions. In contrast, hash functions suffer from the pigeonhole principle (domain larger than range), making collisions inevitable even without truncation.

\textbf{Why is the 0.27\% FNR not exploitable?}
\begin{itemize}
\item \textbf{Random nature of noise:} Noise randomly converts pixels to 0 or 255, and the attacker cannot control it. Unlike structural collisions in hash functions, this FNR arises from pure chance.
\item \textbf{Non-reproducibility:} The noise pattern differs for each image and each execution. Even if an attacker accidentally achieves a collision for a specific block, they cannot reproduce the same pattern for another image (or even another block in the same image).
\item \textbf{Dependence on three key layers:} The TCA output depends on the seed, $K$, and block content. A forged block valid for one parameter combination is useless for another.
\item \textbf{Comparison with structural FNR in hash-based methods:} In shortened hashing, high FNR arises from structural collisions (reduced output space, e.g., 64 bits for an $8\times 8$ block), and an attacker can analytically (or by building a codebook) find two different blocks with the same hash and use this for targeted forgery. In our method, the 0.27\% FNR arises from pure chance and has no exploitable analytical structure.
\end{itemize}

\textbf{Conclusion:} The 0.27\% FNR is a random, non-reproducible event that cannot be used for targeted forgery. This contrasts with shortened hash-based methods, where high FNR arises from structural weaknesses and is exploitable.

\subsection{Extension to Color Images: Results and Analysis}
To demonstrate generalizability to color spaces, 10 standard MATLAB color images (autumn, trees, fabric, flamingos, llama, peacock, hallway, indiancorn, kids, monkey) were resized to $512\times 512$. Using the vertical sandwiching strategy, the images were transformed into a structure of size $1536\times 512$. Parameters were set identically to the grayscale case ($m=n=6$, $K=21.2$, $seed = 123456$), and all 18 attacks were applied.

Results for 17 attacks (excluding salt-and-pepper noise): All attacks, including Gaussian noise, Poisson noise, speckle noise, brightness adjustment, contrast adjustment, median filtering, rotation, ellipse removal, diagonal line addition, JPEG compression, block mean replacement, copy-move, collage (different seed), collage (different K), Photoshop collage, complete pixel change, and single-bit change, were applied to the 10 color images. In all cases, perfect detection (as in Table IV) was achieved, matching the grayscale results.

\begin{table*}[!t]
\centering
\footnotesize
\caption{Results for 17 Attacks on Color Images}
\label{tab:color17}
\setlength{\tabcolsep}{3pt}
\begin{tabular}{|l|c|c|c|c|}
\hline
\textbf{Attack} & \textbf{Parameters} & \textbf{TPR (\%)} & \textbf{FPR (\%)} & \textbf{FNR (\%)} \\
\hline
Gaussian Noise & $\sigma^2 \in \{0.0001, 0.0005, 0.001, 0.005, 0.01, 0.02, 0.05, 0.1\}$ & 100 & 0 & 0 \\
\hline
Poisson Noise & - & 100 & 0 & 0 \\
\hline
Speckle Noise & $d \in \{0.002, 0.02, 0.2\}$ & 100 & 0 & 0 \\
\hline
Brightness Adjustment & $\Delta \in \{1, 10, 50, 100, 149, 200, 243, 254, -254\}$ & 100 & 0 & 0 \\
\hline
Contrast Adjustment & $\gamma \in \{0.1, 0.2, 0.5, 0.8, 0.9\}$ & 100 & 0 & 0 \\
\hline
Median Filtering & $3\times 3$ kernel & 100 & 0 & 0 \\
\hline
Rotation & $1^\circ, 90^\circ, 180^\circ$ (crop/loose) & 100 & 0 & 0 \\
\hline
Ellipse Removal & Random radii (5–15\% of dimensions) & 100 & 0 & 0 \\
\hline
Diagonal Line Addition & Main diagonal & 100 & 0 & 0 \\
\hline
JPEG Compression & $Q \in \{5, 10, 20, 30, 40, 50, 60, 70, 80, 90, 95\}$ & 100 & 0 & 0 \\
\hline
Block Mean Replacement & All blocks & 100 & 0 & 0 \\
\hline
Copy-Move & 1 random block & 100 & 0 & 0 \\
\hline
Collage (Different Seed) & $seed=123456$ vs $123457$ & 100 & 0 & 0 \\
\hline
Collage (Different K) & $K=21.2$ vs $21.1$ & 100 & 0 & 0 \\
\hline
Photoshop Collage & Arbitrary manipulation & 100 & 0 & 0 \\
\hline
Complete Pixel Change & 1 random pixel & 100 & 0 & 0 \\
\hline
Single-Bit Change & Bits 0 to 7 & 100 & 0 & 0 \\
\hline
\end{tabular}
\end{table*}

\textbf{Salt-and-Pepper Noise Results:} The only attack with non-zero FNR. Table V shows the average results over 10 images with standard deviation. Each density was repeated three times.

\begin{table*}[!t]
\centering
\footnotesize
\caption{Salt-and-Pepper Noise Results on 10 Color Images (Mean $\pm$ Std)}
\label{tab:color_sp}
\setlength{\tabcolsep}{3pt}
\begin{tabular}{|c|c|c|c|}
\hline
\textbf{Density ($d$)} & \textbf{TPR (\%)} & \textbf{FPR (\%)} & \textbf{FNR (\%)} \\
\hline
0.0001 & $99.87 \pm 0.41$ & $0 \pm 0$ & $0.13 \pm 0.41$ \\
\hline
0.0005 & $99.77 \pm 0.28$ & $0 \pm 0$ & $0.23 \pm 0.28$ \\
\hline
0.001 & $99.73 \pm 0.26$ & $0 \pm 0$ & $0.27 \pm 0.26$ \\
\hline
0.005 & $99.78 \pm 0.11$ & $0 \pm 0$ & $0.22 \pm 0.11$ \\
\hline
0.01 & $99.81 \pm 0.06$ & $0 \pm 0$ & $0.19 \pm 0.06$ \\
\hline
0.02 & $99.91 \pm 0.05$ & $0 \pm 0$ & $0.09 \pm 0.05$ \\
\hline
0.05 & $99.99 \pm 0.01$ & $0 \pm 0$ & $0.01 \pm 0.01$ \\
\hline
0.1 & $100 \pm 0$ & $0 \pm 0$ & $0 \pm 0$ \\
\hline
\end{tabular}
\end{table*}

FPR is zero at all densities. FNR is near zero at low densities (max 0.27\%) and decreases to zero as density increases. Performance on color images is better than grayscale (for $d=0.001$, FNR dropped from 0.41\% to 0.27\%); the reason is the threefold increase in pixels affected by noise (three channels), reducing the probability of pure coincidence.

\textbf{Visual Quality and Execution Time:} Average $PSNR = 51.14$ dB and $SSIM = 0.9990$, indicating excellent visual quality and imperceptibility. Embedding time = 2.61 s and extraction time = 2.52 s. Compared to grayscale (1.61 s), the time increased by about 1.62 times (not 3 times) due to maintaining the number of blocks in vertical sandwiching—although block height increased from $6\times 6$ to $18\times 6$, the number of blocks (7225) remained unchanged. In contrast, if the three color channels were processed independently, the number of blocks would triple and time would approximately triple. This is a structural advantage of vertical sandwiching.

To visually illustrate the effectiveness of the proposed method for color images, we present a practical example in Fig. 4 using the 'peacock' image from the standard MATLAB image set. In this example, a Photoshop collage attack is applied to the watermarked color image. The tampered region is clearly marked with a red circle in the attacked image (Fig. 4(b)). The extraction algorithm operates on the vertically sandwiched block image (Fig. 4(c)), which stacks the R, G, and B channels of each block, resulting in a height three times that of the original. The resulting tamper map in the sandwiched domain is shown in Fig. 4(d), where the tampered blocks are precisely detected across the three channel stacks. Finally, the tamper map is resized to the original image dimensions, accurately localizing the tampered region as shown in Fig. 4(e). This example demonstrates the method's capability to detect and localize sophisticated attacks such as collage in color images.

\begin{figure*}[!t]
\centering
\includegraphics[width=\textwidth]{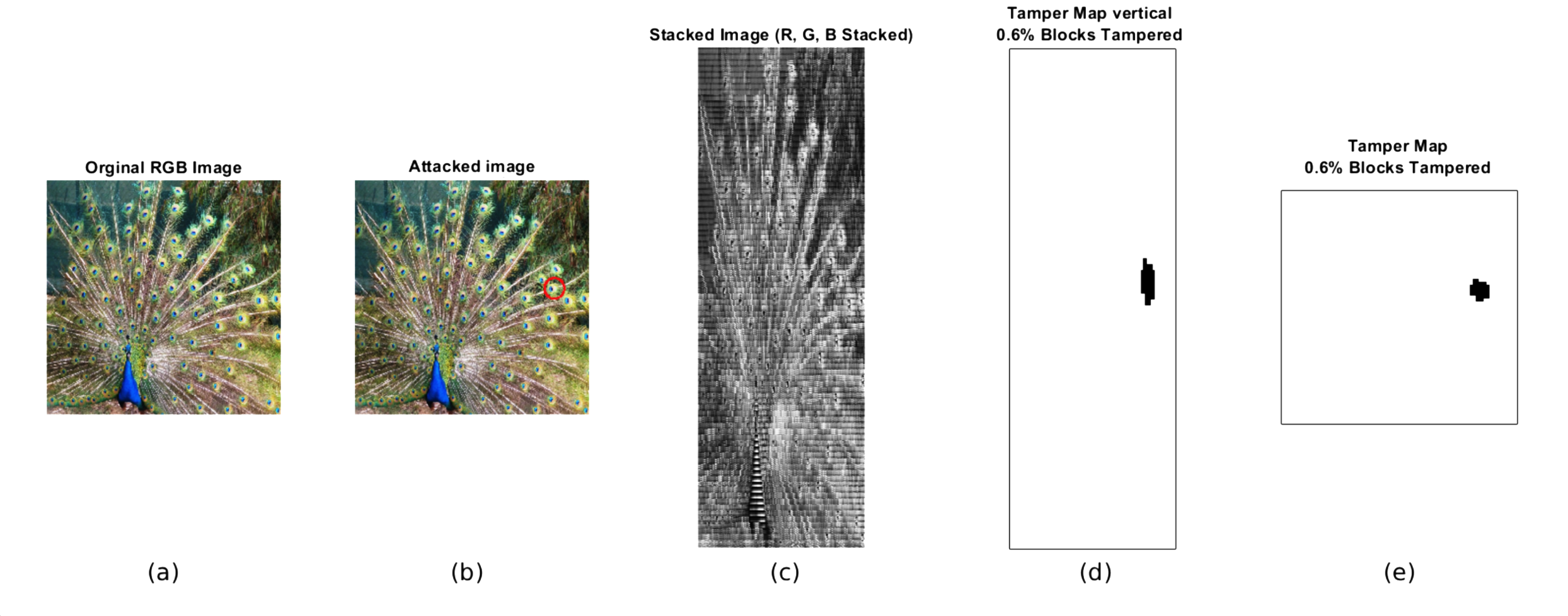}
\caption{Example of tamper localization for a color image under a Photoshop collage attack using the 'peacock' image from the MATLAB image set. (a) Original RGB image. (b) Attacked image; the tampered area (collage) is highlighted by a red circle. (c) Vertically sandwiched block image constructed by stacking the R, G, and B channels of each block of the attacked image. (d) Tamper map generated by the extraction algorithm in the sandwiched domain; note that the height is three times the original image height due to the vertical stacking of the three color channels. (e) Final tamper map resized to the original image dimensions (H$\times$W), accurately localizing the tampered region.}
\label{fig:color_tamper_example}
\end{figure*}

\textbf{Summary:} (1) Extension to color without changing the core algorithm; (2) perfect detection for 17 attacks; (3) negligible FNR (max 0.27\%) for salt-and-pepper noise, decreasing to zero with increasing density—better performance than grayscale; (4) excellent visual quality and reasonable execution time. These features distinguish the method from many existing approaches designed only for 8-bit grayscale images.

\subsection{Extension to High Bit-Depth Images}
To validate scalability to 12-bit and 16-bit depths, synthetic $512\times 512$ images with random pixel values were generated in MATLAB. The reason for using synthetic images was the lack of access to a standard dataset of high bit-depth medical images. However, since TCA uses local image features (edges and Delaunay triangulation) to generate the permutation map, and the shuffling process itself is a purely mathematical and deterministic procedure, the results readily generalize to real high bit-depth images (such as CT scans or MRI).

Parameters were set identically to the 8-bit case, with adjustments to modular operations and noise generation for the bit-depth (Section 4.12). Since the algorithm follows the same deterministic structure in higher depths, performance against previous attacks remains unchanged. Therefore, only two attacks with potential for different behavior were evaluated: (1) single-bit change and (2) salt-and-pepper noise (the only attack with non-zero FNR in 8-bit).

Results are shown in Table VI.

\begin{table}[!t]
\centering
\footnotesize
\caption{Performance on Synthetic High Bit-Depth Images ($512\times 512$)}
\label{tab:highbit}
\setlength{\tabcolsep}{3pt}
\begin{tabular}{|c|c|c|c|c|}
\hline
\textbf{Bit-Depth} & \textbf{Attack} & \textbf{TPR (\%)} & \textbf{FPR (\%)} & \textbf{FNR (\%)} \\
\hline
12-bit & Single-Bit Change & 100 & 0 & 0 \\
\hline
12-bit & Salt-and-Pepper Noise (0.05) & 99.97 & 0 & 0.03 \\
\hline
16-bit & Single-Bit Change & 100 & 0 & 0 \\
\hline
16-bit & Salt-and-Pepper Noise (0.05) & 100 & 0 & 0 \\
\hline
\end{tabular}
\end{table}

\textbf{Analysis of Results}
\begin{itemize}
\item \textbf{Visual Quality:} With 5 repetitions, the average PSNR for 12-bit images was 75.25 dB, and for 16-bit images, it was 99.33 dB. The high values indicate imperceptibility and preservation of the full dynamic range. The increase in PSNR with bit-depth is consistent with theoretical expectations.
\item \textbf{Execution Time:} \textbf{Execution Time:} With 5 repetitions, the average embedding and extraction times for 12-bit images were 1.89 s and 1.88 s, respectively, and for 16-bit images, they were 2.05 s and 1.97 s, respectively. The increase in time at higher bit-depths is due to the larger TCA bit-matrix (from $18\times 24$ for 12-bit to $24\times 24$ for 16-bit).
\item \textbf{Sensitivity to Single-Bit Changes:} At both depths, the method detected a random single-bit change with 100\% accuracy ($TPR=100\%$, $FNR=0\%$). Five repetitions on 12-bit with different bit positions and locations all resulted in perfect detection. This shows that single-bit sensitivity is preserved independently of bit-depth and without normalization—unlike methods such as~\cite{Bouarroudj2025}, which lose subtle changes due to normalization to 8 bits.
\item \textbf{Salt-and-Pepper Noise:} At 16-bit, FNR under noise (density 0.05) dropped to zero, whereas for 8-bit the value was 0.15\% (for density 0.05). At 12-bit, FNR also dropped to 0.03\%. This improvement is attributed to the larger bit-matrix size at higher depths ($24\times 24$ for 16-bit, $18\times 24$ for 12-bit, versus $12\times 24$ for 8-bit). TCA operates on a richer bit representation, and the 36-bit watermark selected from the center becomes more resistant to random collisions without compromising fragility.
\end{itemize}

\textbf{Conclusion:} The method seamlessly scales to 12-bit and 16-bit depths. Unlike methods that require normalization to 8 bits, our method involves all image bits in tamper detection. This feature, combined with 100\% sensitivity to single-bit changes and high PSNR, makes the method suitable for sensitive applications such as medical images (CT scans, MRI), where preserving subtle diagnostic information is critical.

\subsection{Experimental Verification of Generalizability to Arbitrary Dimensions and User-Selectable Block Size}
To simultaneously verify two main claims—(1) support for arbitrary dimensions without padding, and (2) user freedom in selecting block size—a challenging experiment was designed. An image of size $610\times 1027$ (which is divisible by no regular block size) was selected with a non-square block size of $7\times 4$. Results are summarized in Table VII.

\begin{table}[!t]
\centering
\footnotesize
\caption{Results of Experiment with Arbitrary Dimensions and Non-Square Block}
\label{tab:arbitrary}
\setlength{\tabcolsep}{3pt}
\begin{tabular}{|l|c|}
\hline
\textbf{Parameter} & \textbf{Value} \\
\hline
Image Dimensions & $610 \times 1027$ \\
\hline
Block Size & $7 \times 4$ \\
\hline
Number of Blocks & 22272 \\
\hline
PSNR (dB) & 51.15 \\
\hline
Embedding Time (avg. 5 runs) & 5.78 s \\
\hline
Extraction Time (avg. 5 runs) & 5.47 s \\
\hline
Clean Image Detection & $\checkmark$ (FPR=0\%) \\
\hline
Single-Bit Change Detection (5 runs) & $\checkmark$ (TPR=100\%, FNR=0\%) \\
\hline
Duplicate Watermarks & 1 case (out of 22272 blocks) \\
\hline
\end{tabular}
\end{table}

The algorithm operated without error. $PSNR=51.15$ dB is comparable to standard results. The clean image was correctly verified (FPR=0\%). Single-bit changes in 5 runs were all successfully detected ($TPR=100\%$, $FNR=0\%$). Out of 22272 blocks, only 1 duplicate watermark was observed, indicating exceptionally high distinctiveness.

\subsection{Analysis of Block Size Impact on Efficiency and Security}
To investigate the impact of block size on execution time and watermark uniqueness, the method was run on a fixed $512\times 512$ image (Baboon grayscale) with various block sizes. Results are summarized in Table VIII.

\begin{table*}[!t]
\centering
\footnotesize
\caption{Performance Comparison at Different Block Sizes ($512\times 512$ Image)}
\label{tab:blocksize}
\setlength{\tabcolsep}{3pt}
\begin{tabular}{|c|c|c|c|c|c|}
\hline
\textbf{Block Size} & \textbf{Number of Blocks} & \textbf{Duplicate Watermarks (Pairs)} & \textbf{Embedding Time (s)} & \textbf{Extraction Time (s)} & \textbf{PSNR (dB)} \\
\hline
$3\times 3$ & 28,900 & 798,842 & 5.91 & 7.40 & 51.14 \\
\hline
$4\times 4$ & 16,384 & 2,076 & 3.69 & 3.50 & 51.13 \\
\hline
$5\times 5$ & 10,404 & 1 & 2.44 & 2.38 & 51.13 \\
\hline
$6\times 6$ & 7,225 & 0 & 1.61 & 1.63 & 51.14 \\
\hline
$8\times 8$ & 4,096 & 0 & 1.19 & 1.18 & 51.15 \\
\hline
$16\times 16$ & 1,024 & 0 & 0.71 & 0.84 & 51.14 \\
\hline
$32\times 32$ & 256 & 0 & 0.60 & 0.66 & 51.14 \\
\hline
\end{tabular}
\end{table*}

The number of duplicate pairs indicates pairwise collisions between watermarks. A value greater than the number of blocks (e.g., $3\times 3$) indicates severe repetition, as each group of identical watermarks creates multiple duplicate pairs.

\textbf{Analysis of Results:}
\begin{itemize}
\item \textbf{Impact on Execution Time:} As block size increases, the number of blocks decreases, and execution time decreases significantly (from 5.91 s for $3\times 3$ to 0.60 s for $32\times 32$), due to fewer TCA executions.
\textbf{Impact on Execution Time:} As block size increases, the number of blocks decreases, and execution time decreases significantly (from 5.91 s for $3\times 3$ to 0.60 s for $32\times 32$), due to fewer TCA executions.
\item \textbf{Impact on Watermark Uniqueness (Security):} A direct relationship exists between block size and the number of duplicate watermarks. At $3\times 3$, the limited output space (9 bits) results in 798,842 duplicate pairs (severe collisions). At $4\times 4$, 2,076 pairs; at $5\times 5$, 1 duplicate pair is observed—even a single collision poses a threat to block relocation attacks. At $6\times 6$ and larger, no duplicate watermarks were observed (complete uniqueness). This feature makes the method fully resistant to codebook (VQ) and block relocation attacks. The reason is the increased output space with block size; for $6\times 6$, the 36-bit watermark has $2^{36}$ (approximately 68 billion) possible states, making collision finding impossible.
\item \textbf{Impact on Visual Quality:} PSNR is approximately 51.14 dB for all block sizes and independent of block size.
\end{itemize}

\textbf{Final Conclusion:} Block size $6\times 6$ was selected as the minimum secure size. Reasons: (1) complete security (no duplicate watermarks and resistance to VQ/collage); (2) reasonable localization accuracy (36 pixels vs. larger blocks); (3) reasonable efficiency (1.61 s embedding); (4) excellent visual quality ($PSNR=51.14$ dB).

It should be noted that smaller blocks (e.g., $3\times 3$ or $4\times 4$) offer higher localization accuracy but lack the necessary security due to severe watermark collisions. On the other hand, larger blocks (e.g., $16\times 16$ or $32\times 32$) provide complete security but reduce localization accuracy. Therefore, $6\times 6$ was chosen as the optimal trade-off point.

All attack experiments in this paper (Sections 5.3–5.8) were performed with this configuration, and the results demonstrate excellent performance with FPR=0 and near-zero FNR.


\section{Limitations and Future Work}

\subsection{Limitations}

\begin{itemize}
\item \textbf{Computational complexity for real-time applications:} Although the embedding time of 1.61 s and extraction time of 1.63 s for $512\times 512$ images are suitable for near real-time and all offline applications such as medical image archiving and digital forensics, further optimization is still required for strict real-time applications (e.g., live video streaming at 30 fps). The main sources of overhead remain Canny edge detection and Delaunay triangulation within the TCA core.

\item \textbf{Block-level localization granularity:} The method localizes tampering at the block level ($6\times 6$, 36 pixels). While acceptable for many applications, highly localized attacks affecting only a few pixels are not pinpointed at the pixel level; the entire block is marked as tampered.
\end{itemize}

\subsection{Future Work}

\begin{itemize}
\item \textbf{Real-time optimization:} Preliminary investigations indicate that replacing Canny with Sobel edge detection can reduce execution time for $6\times 6$ blocks from 1.61 s to approximately 0.49 s (a reduction of $>69\%$), presenting a promising path toward real-time optimization. However, the impact of this change on the avalanche effect, watermark uniqueness, and single-bit sensitivity requires comprehensive verification. Additionally, GPU/FPGA acceleration and rewriting the core TCA routines in C/C++ with SIMD optimizations could significantly reduce execution time.

\item \textbf{Pixel-level localization:} Extending the scheme to single-pixel accuracy through hierarchical watermarking (combining $8\times 8$, $4\times 4$, and $2\times 2$ blocks) or using overlapping blocks with majority voting could improve localization accuracy without compromising security. These approaches could achieve a better trade-off between accuracy and security.

\item \textbf{Extension to video and 3D data:} Generalizing the TCA concept to video sequences (treating frames as a third dimension) and volumetric medical images (e.g., CT or MRI) using 3D Delaunay tetrahedralization instead of 2D triangulation is another promising direction.

\item \textbf{Native color image processing:} Developing a version of TCA that operates natively on color images without requiring vertical sandwiching, using vector-based edge detection (e.g., color gradient) directly on RGB or YCbCr channels, could improve efficiency.
\end{itemize}

\section{Conclusion}

In this paper, we presented a novel fragile watermarking method based on Triangular Content-Aware Permutation (TCA) at the bit-plane level. Unlike traditional methods relying on fixed mathematical transforms, our approach leverages the intrinsic avalanche effect of TCA to generate a content-dependent and highly sensitive watermark. The core innovation lies in transferring TCA from the pixel domain to the bit-plane level, enabling single-bit sensitivity while preserving the pure permutation nature.

The method addresses eight fundamental research gaps: (G1) guaranteed zero FPR; (G2) resistance to VQ and collage; (G3) full fragility against JPEG; (G4) single-bit sensitivity; (G5) immunity to rounding errors; (G6) support for high bit-depths without normalization; (G7) user-selectable block size; and (G8) arbitrary dimensions without padding. To the best of our knowledge, no prior work simultaneously addresses all eight gaps.

The method comprises three independent security layers: (a) seed-based noise masking for spatial binding; (b) CTR-style multipliers for block order binding; and (c) bit-level TCA scrambling for the avalanche effect. The ``remainder merging'' strategy eliminates dimension constraints, allowing the algorithm to run without padding on images of any size. The method naturally extends to color images via vertical sandwiching, which preserves inter-channel dependencies and increases time by only 1.62 times (not 3 times). Extension to high bit-depths (12 and 16 bits) is achieved with three simple parameter adjustments and no normalization.

Experimental evaluation on 50 grayscale and 10 color images under 18 attacks shows that the method achieves perfect detection ($TPR=100\%$, $FPR=0\%$, $FNR=0\%$) for 17 attacks. Only for salt-and-pepper noise is there a negligible FNR of 0.27\% for grayscale and 0.14\% for color images. Vertical sandwiching reduces FNR by approximately 0.13\% compared to grayscale. Visual transparency is confirmed by average PSNR of 51.14 dB (8-bit), 75.25 dB (12-bit), and 99.33 dB (16-bit). Embedding and extraction times for grayscale are 1.61 s and 1.63 s, respectively. The algorithm achieves 100\% accuracy against collage, VQ, copy-move, JPEG compression (quality 5--95), and geometric attacks.

Pairwise distinctiveness testing on over 72,820 blocks confirmed zero duplicate watermarks, proving that VQ and collage attacks are impossible. The negligible 0.27\% FNR for salt-and-pepper noise arises from pure chance, not structural collisions, making it unexploitable for targeted forgery---in contrast to shortened hash-based methods, where high FNR arises from exploitable structural weaknesses.

The method provides a secure and flexible solution for sensitive applications such as digital forensics, medical image archiving, and legal document authentication. By replacing static mathematical formulas with dynamic content-dependent geometry, our method achieves both high security and practical efficiency. Future work will focus on real-time optimization (GPU acceleration, low-level language implementation), pixel-level localization (hierarchical or overlapping blocks), and extension to video and 3D data.

\section*{Acknowledgments}
The authors would like to thank the anonymous reviewers for their valuable comments and suggestions.

\section*{Author Contributions}
\textbf{Zahra Ghoraeian:} Conceptualization, Methodology, Software, Data Curation, Formal analysis, Visualization, Writing--original draft.\\
\textbf{Mohammad-Reza Sadeghi:} Supervision, Validation, Resources, Writing--review \& editing, Project administration.\\
\textbf{Samaneh Mashhadi:} Methodology, Validation, Formal analysis, Visualization, Writing--review \& editing.

\section*{Funding}
This research received no specific grant from any funding agency in the public, commercial, or not-for-profit sectors.

\section*{Conflict of Interest}
The authors declare that they have no known competing financial interests or personal relationships that could have appeared to influence the work reported in this paper.

\section*{Data Availability Statement}
The complete MATLAB source code, including embedding and extraction scripts 
for grayscale, RGB, and high bit-depth images, along with all test images 
and sample datasets supporting the findings of this study, are publicly 
available on Zenodo at \url{ https://doi.org/10.5281/zenodo.21036130}.
\bibliographystyle{IEEEtran}
\bibliography{references}

\end{document}